\documentclass[fleqn,usenatbib]{mnras}

\usepackage{newtxtext,newtxmath}

\usepackage[T1]{fontenc}

\DeclareRobustCommand{\VAN}[3]{#2}
\let\VANthebibliography\thebibliography
\def\thebibliography{\DeclareRobustCommand{\VAN}[3]{##3}\VANthebibliography}

\usepackage{graphicx}	
\usepackage{amsmath}	
\usepackage{multirow}
\usepackage{lscape}
\usepackage{float}

\usepackage{xcolor}
\usepackage{rotating} 
\usepackage{tablefootnote}
\usepackage{hyperref}
\usepackage{bm}
\usepackage{cite}
\usepackage[capitalise]{cleveref}

\usepackage{tocloft}

\DeclareRobustCommand{\svdots}{
  \vbox{%
    \baselineskip=0.33333\normalbaselineskip
    \lineskiplimit=0pt
    \hbox{.}\hbox{.}\hbox{.}%
    \kern-0.2\baselineskip
  }%
}

\newcommand{\Mpch}{\ensuremath{h^{-1}\, \text{Mpc}}}

\title{Symbolically regressing dark matter halo profiles using weak lensing}

\author[A. Martín et al.]{
Alicia Martín,$^{1}$\thanks{E-mail: alicia.martin@physics.ox.ac.uk}
Tariq Yasin,$^{1}$
Deaglan J. Bartlett,$^{1,2}$
Harry Desmond$^{3}$
and Pedro G. Ferreira$^{1}$
\\
$^{1}$Astrophysics, University of Oxford, Oxford, OX1 3RH, United Kingdom\\
$^{2}$CNRS \& Sorbonne Université, Institut d’Astrophysique de Paris (IAP), UMR 7095, 98 bis bd Arago, F-75014 Paris, France\\
$^{3}$Institute of Cosmology \& Gravitation, University of Portsmouth, Portsmouth, PO1 3FX, United Kingdom
}

\date{Accepted XXX. Received YYY; in original form ZZZ}

\pubyear{\the\year{}}

\begin{document}
\label{firstpage}
\pagerange{\pageref{firstpage}--\pageref{lastpage}}
\maketitle

\begin{abstract}
The structure of dark matter haloes is often described by radial density profiles motivated by cosmological simulations. These are typically assumed to have a fixed functional form (e.g. NFW), with some free parameters that can be constrained with observations. However, relying on simulations has the disadvantage that the resulting profiles depend on the dark matter model and the baryonic physics implementation, which are highly uncertain. Instead, we present a method to constrain halo density profiles directly from observations. This is done using a symbolic regression algorithm called Exhaustive Symbolic Regression (ESR). ESR searches for the optimal analytic expression to fit data, combining both accuracy and simplicity. We apply ESR to a sample of $149$ galaxy clusters from the HSC-XXL survey to identify which functional forms perform best across the entire sample of clusters. We identify density profiles that statistically outperform NFW under a minimum-description-length criterion. 
Within the radial range probed by the weak-lensing data ($R \sim 0.3 - 3$ h$^{-1}$ Mpc), the highest-ranked ESR profiles exhibit shallow inner behaviour and a maximum in the density profile. However, the inner slope itself remains weakly constrained due to limited signal at small radii. As a practical application, we show how the best-fitting ESR models can be used to obtain enclosed mass estimates. We find masses that are, on average, higher than those derived using NFW, highlighting a source of potential bias when assuming the wrong density profile. These results have important knock-on effects for analyses that utilise clusters, for example cosmological constraints on $\sigma_8$ and $\Omega_\text{m}$ from cluster abundance and clustering. Beyond the HSC dataset, the method is readily applicable to any data constraining the dark matter distribution in galaxies and galaxy clusters, such as other weak lensing surveys, galactic rotation curves, or complementary probes. 
\end{abstract}




\section{Introduction}


In the $\Lambda$CDM picture, structure grows hierarchically. Small primordial fluctuations are amplified by gravity, first forming small haloes and then, through accretion and mergers, building ever larger systems that eventually virialise. At the top of this hierarchy sit galaxy clusters. They are the most massive bound objects in the Universe and are strongly dark-matter dominated \citep{allen_2011}. As such, they are natural laboratories for testing the properties of dark matter and, more broadly, for probing possible departures from standard gravity \citep{kravtson_2012}.

Consequently, clusters are powerful cosmological probes. Their abundance, when combined with robust mass calibration, places tight constraints on parameters such as the matter density, $\Omega_{\rm m}$, and the amplitude of fluctuations, $\sigma_8$ \citep{vikhlinin_2009, abbott_2020_des}. Beyond abundance statistics, their internal mass profiles encode their specific assembly histories. These imprints are visible in structural features like the splashback radius \citep{diemer_2014, more_2015} and in the scaling relations that connect baryons to the total gravitational field \citep{Chan_2020, tian2020radial, Tian_2024, Mistele_2025}. Therefore, characterising the density profiles of clusters informs both the astrophysics of structure formation and the underlying cosmological model.


For decades, large-volume cosmological simulations have been the primary tool for studying the structure of dark matter haloes. Early $N$-body simulations consistently showed that haloes exhibit a steep power-law increase in density toward the centre, known as a ``cusp'' \citep{dubinski1991structure, navarro1997universal}. This initial finding appeared to be universal across different simulations and consistent over a wide range of halo masses, concentrations, and cosmological models, and was captured by the Navarro-Frenk-White (NFW) profile \citep{navarro1997universal}
\begin{equation}
    \rho(r) = \frac{\rho_0}{r/r_s( 1 + r/r_s)^2},
    \label{eq:NFW}
\end{equation}
where $\rho_0$ is a characteristic density and $r_s$ a scale radius. The NFW profile follows a double power law that transitions from $\rho(r) \propto r^{-1}$ at small radii to $\rho(r) \propto r^{-3}$ at large radii.

However, the notion of a universal NFW profile was challenged by later, higher-resolution dark matter-only simulations, which revealed more diversity in inner density slopes. For example, the Aquarius \citep{navarro_inner} and Via Lactea II \citep{diemand2007formation} simulations found that the logarithmic slope becomes progressively shallower than $-1$ toward the centre. In contrast, \citet{moore1998resolving} reported a steeper inner slope of $\rho(r) \sim r^{-1.4}$. More recent high-resolution simulations appear to support this steeper behaviour, finding $\rho(r) \sim r^{-1.5}$
\citep{delos2019predicting, delos2025cusp}, which they refer to as a ``prompt cusp''. This wider class of generalised NFW models suggests that factors such as numerical convergence, initial conditions or analysis techniques may still influence the inferred inner slope. 

This disagreement has also motivated the development of alternative, more flexible density models. In particular, the Einasto profile \citep{einasto}, which assumes a power-law form for the logarithmic density slope (${\rm d}\ln\rho/{\rm d}\ln r \propto r^{\alpha}$) rather than for the density itself, was advocated by \citet{navarro_inner} and \citet{Merritt_2005}. Using high-resolution $N$-body simulations these works and subsequent studies \citep{Merritt_2006, Gao_2008, navarro2010diversity, Ludlow_2013} showed that the Einasto model provides a better fit to halo density profiles than NFW, especially in the inner regions.

Beyond collisionless dynamics, hydrodynamical simulations further broaden the possibilities: gas cooling and the assembly of the brightest cluster galaxy (BCG) can contract the dark matter and steepen inner slopes \citep{blumenthal_1986, laporte2012shallow}, whereas energetic feedback from active galactic nuclei can counteract or even flatten the central profile \citep{martizzi2013cusp, peirani2017density}. These processes occur on scales close to current resolution limits and are modelled with subgrid prescriptions, so inferred inner slopes can depend on both numerical set-ups and baryonic physics choices \citep{crain2015eagle, sorini2025impact}.

Another limitation of simulation-based profiles is that they are phenomenological: they are simple, low-parameter functions calibrated to match the output of collisionless $N$-body simulations, rather than solutions derived from a fundamental theory. In recent years, however, progress has been made towards a first-principles understanding. For example, \citet{pontzen2013conserved} developed an analytical framework to study the phase space distribution of cold dark matter particles based on entropy considerations. Their framework reproduces the shallower inner cusp and steep outer decline, in qualitative agreement with simulated haloes. More recently, \citet{banik2025collisionless} showed that the NFW profile might emerge as an attractor solution to a self-consistent theory for collisionless relaxation. While these works offer some insight on why such profiles might appear in cold dark matter simulations, their theoretical motivation is still incomplete.

More significantly, observational data often challenge the universality of the NFW profile, particularly regarding its cuspy inner structure. On galactic scales, observations of dwarf and low-surface-brightness galaxies suggest a flatter inner density shape, even reaching a constant density region known as a ``core'' \citep{Flores_cores, moore_1994, deblock_lsb}. On cluster scales, the picture is more complex. While stacked lensing generally favours cuspy profiles in the $0.1$–$2\,\mathrm{Mpc}$ regime \citep{umetsu2016}, detailed analyses combining strong lensing with stellar kinematics sometimes prefer shallower slopes within the inner $10$–$50\,\mathrm{kpc}$ \citep{sand2002dark, newman2011dark}.

Interpreting these potential tensions is difficult. They may arise from observational systematics within the standard $\Lambda$CDM framework, such as triaxiality, miscentering, or non-thermal pressure support \citep{CorlessKing2007, JohnstonEtAl2007, LauKravtsovNagai2009}. Alternatively, they may point to new physics. For example, allowing for self-interacting dark matter (SIDM) can naturally produce core-like profiles that deviate from the standard collisionless prediction \citep{spergel2000observational, Rocha_2013, Ragagnin_2024, zeng2025tillcorecollapsesevolution}. 

These uncertainties highlight the risks of imposing fixed density templates on observational data. When a single parametric form is assumed, any deviation in the underlying density profile can propagate directly into biased mass or concentration estimates. For example, in cluster mass calibration, forcing an NFW profile onto a halo that has been modified by baryonic feedback or is in a non-equilibrium state can skew the inferred total mass. Such model-dependent systematics are a leading explanation for the tension between cluster weak-lensing masses and those inferred from \emph{Planck} CMB observations \citep{von_der_Linden_2014, Hoekstra_2015, Blanchard_2021}. These issues emphasise the need for approaches that allow the data to determine the form of the density profile.

To address this, we propose a data-driven alternative: deriving density profiles directly from observations using Symbolic Regression \citep[SR; see][for a recent review]{Kronberger_2024}. SR is a supervised machine learning method which tries to find the best analytical function to describe a given dataset. It searches the mathematical space of functions, fitting them to the data to identify the most accurate models. Importantly, it balances accuracy with simplicity, helping to avoid over-fitting and returning functions that are both interpretable and generalisable.

In particular, we use a novel SR algorithm called Exhaustive Symbolic Regression (hereafter ESR;~\citealt{bartlett_exhaustive_2024}). Unlike other heuristic or evolutionary approaches, ESR explores all possible mathematical models up to a specified functional complexity (defined to be equal to the number of nodes in the tree representation of a function), ensuring that the best model to describe the data is always evaluated provided it is within that complexity limit. ESR combines simplicity and accuracy by using the Minimum Description Length (MDL) principle, which ranks functions according to the number of bits of information required to encode the data with the help of the function.

In \citealt{martín2025constraining}, we tested the approach on mock weak lensing excess surface density (ESD) data of synthetic clusters with NFW profiles. Motivated by real data, we assigned each ESD data point a constant fractional uncertainty and varied this uncertainty and the number of clusters to probe how data precision and sample size affect model selection. For fractional errors around 5 per cent, we showed that ESR recovers the NFW profile even for a low number of samples. At higher uncertainties representative of current surveys, simpler functions are favoured over NFW, although NFW typically remains within the top $\sim 10$ models.

In this paper, we extend this analysis to real observations by applying ESR to stacked weak lensing (WL) measurements from a sample of $149$ galaxy clusters observed by the Hyper Suprime-Cam (HSC) survey~\citep{adami2018xxl}. Our goal is to find the analytical profiles that best reproduce the ESDs of this cluster sample and test whether NFW remains among the highest-ranked profiles, or whether the observational data favour alternative functional forms. We can further use the ESR-derived profiles to estimate properties of the clusters, such as the enclosed mass. By averaging over all plausible models, this mass estimate is insensitive to profile assumptions and therefore more robust than those obtained from traditional NFW fitting, which in general have an unknown systematic errors associated with the fixed profile.

While this work focuses on clusters, the methodology is readily applicable to the full range of data constraining the dark matter distribution around galaxies, groups, and clusters, as well as many other astrophysical observables.

The paper is structured as follows. In \cref{section:data}, we give a description of the WL shear data used in this study and the calculation of ESD profiles from it. \cref{section:ESR} summarises the ESR algorithm, while \cref{section:Methods} describes how to use ESR to model the ESD profiles and the specific settings used to optimise parameters and estimate their uncertainties. The results are presented in \cref{section:results}, followed by the discussion in \cref{section:discussion} and conclusion in \cref{section:conclusions}.

Throughout this paper, we assume a spatially flat $\Lambda$CDM cosmology with  $\Omega_{\rm m} = 0.28$, $\Omega_{\Lambda} = 0.72$ and a value of the Hubble constant of $H_0 = 100 h$ km s$^{-1}$ Mpc$^{-1}$ with $h = 0.7$ following the HSC$\times$XXL weak-lensing analysis \citep{umetsu_cluster-galaxy_2020}. We use $\log$ to denote the natural logarithm.

\section{Weak lensing data}
\label{section:data}

\subsection{Cluster and Lensing Catalogues}

In this work, we largely follow the cluster selection and WL data processing of \citet{umetsu_weak_2020}. Our analysis is based on a sample of X-ray selected galaxy groups and clusters from the XMM-XXL survey. We use spectroscopically confirmed systems from the XXL DR2 catalogue presented in \citet{adami2018xxl, mandelbaum2018first, medezinski2018planck}.

WL measurements for these systems are sourced from the first-year data release of the Hyper Suprime-Cam (HSC) Subaru Strategic Program survey \citep[HSC-SSP;][]{aihara2018}. The HSC-SSP survey provides deep optical imaging in five broad bands ($grizy$). Galaxy shape measurements for this catalogue were derived from the co-added $i$-band images using the re-Gaussianisation method \citep{hirata2003}. This $i$-band imaging was performed under exceptional seeing conditions (median FWHM $\simeq 0.6''$). We use the shear catalogue produced for the ``Arcturus'' field, which covers $137$ deg$^2$ over $6$ sky patches. After masking bright stars and other artefacts, this area is reduced to $29.5$ deg$^2$.

We select XXL clusters that fall within the HSC-SSP footprint, resulting in a total overlapping sky area of $21.4$ deg$^2$. Following the criteria in \citet{mandelbaum2018weak}, which requires clusters to be within a comoving separation of $R_{\text{min}} = 0.3 \, \Mpch$ of the lensing data, our final sample consists of $149$ spectroscopically confirmed XXL clusters. The background source galaxies in this matched field have a weighted number density of $\sim 22.1$ galaxies arcmin$^{-2}$ and a mean redshift of $0.82$~\citep{miyatake2019weak}.

\subsection{Shear-to-ESD estimator}


Weak gravitational lensing induces coherent distortions in the shapes of background galaxies as a result of the deflection of light by foreground mass. This effect is sensitive to both baryonic and dark matter and gives a direct probe of the projected mass distribution of a lens. The shape distortion is quantified by the shear, $\gamma$, which can be decomposed into the tangential component $\gamma_+$ and the $45^\circ$-rotated component $\gamma_{\times}$. The tangential shear $\gamma_+$ is directly related to the azimuthally averaged surface mass density $\Sigma(R)$ through~\citep{Bartelmann_2001}
\begin{equation}
    \label{eq:ESD_definition}
   \gamma_{+} = \frac{\langle\Sigma (<R)\rangle - \Sigma(R)}{\Sigma_{\text{cr}}} \equiv \frac{\Delta \Sigma(R)}{\Sigma_{\text{cr}}}, 
\end{equation}
where $R$ is the projected radius from the lens centre,  $\Delta \Sigma(R)$ is the ESD and $\langle\Sigma (<R)\rangle$ is the average surface density within $R$, given by
\begin{equation}
    \langle\Sigma (< R)\rangle = \frac{2}{R^2}\int_{0}^{R}{\Sigma(R^\prime ) R^\prime {\rm d} R^\prime }.
\end{equation}
The critical surface density $\Sigma_{\text{cr}}$ is
\begin{equation}
     \Sigma_{\text{cr}} = \frac{c^2}{4 \pi G}\frac{D_{\rm A}(z_s)}{(1 + z_{\rm l})^2 D_{\rm A}(z_{\rm l}) D_A(z_{\rm l},z_{\rm s})},
\end{equation}
where $c$ is the speed of light, $G$ is the gravitational constant, $z_{\rm l}$ is the lens redshift, $z_{\rm s}$ is the source redshift, and $D_{\rm A}$ is the angular diameter distance. The factor $(1+z_{\rm l})^2$ converts to comoving surface mass density.

To estimate the ESD, we stack the WL signals of background galaxies centred on the X-ray peak position of each cluster. The ESD, $\Delta \Sigma (R)$, is then measured as a function of comoving cluster-centric radius $R$, using $N=9$ radial bins of equal logarithmic spacing from $R_{\text{min}} = 0.3 \, \Mpch$ to $R_{\text{max}} = 3 \, \Mpch$ \citep{medezinski2018planck, miyatake2019weak}. 

The ensemble ESD estimator for a radial bin $i$ is given by

\begin{equation} 
\Delta \Sigma (R_i) = \frac{1}{2\mathcal{R}(R_i)} \frac{\sum_{l,s \in i} w_{ls} e_{+, ls} [\langle \Sigma_{\text{cr}, ls} \rangle]^{-1}}{[1 + K(R_i)] \sum_{l,s \in i} w_{ls}}, \label{eq:ESD}
\end{equation}
where the sum is taken over all lens-source pairs ($l,s$) where the source galaxy $s$ lies within the $i$-th radial bin relative to the lens $l$. Here, $e_{+, ls}$ is the tangential ellipticity of the source galaxy.

The term $\langle \Sigma_{\text{cr}, ls} \rangle^{-1}$ is the lensing strength for each lens-source pair, computed by averaging over the full photometric redshift (photo-z) probability distribution function (PDF) of the source galaxy, $P_s(z)$. The statistical weight, $w_{ls}$, is defined as:
\begin{equation}
w_{ls}=(\langle\Sigma_{cr,ls}^{-1}\rangle)^{2}\frac{1}{\sigma_{e,s}^{2}+e_{{\rm rms},s}^{2}},
\end{equation}
where $\sigma_{e,s}$ is the shape measurement uncertainty and $e_{{\rm rms},s}$ is the root mean squared ellipticity (or shape noise) per component.

This estimator incorporates corrections for two systematic effects: multiplicative bias and shear responsivity. The multiplicative correction factor, denoted as $1+K$, compensates for any systematic under- or over-estimation of the shear magnitude. This bias arises mainly from imperfect modelling of the point-spread function (PSF) and other measurement effects~\citep{kaiser1994method}. In HSC, it is mitigated using a Gaussianiation technique~\citep{mandelbaum2018first}, but a small residual bias remains. For an ensemble of galaxies, this residual is incorporated through a scale-dependent correction factor~\citep{murray_measuring_2022}:
\begin{equation}
1 + K(R_i) = \frac{\sum_{l,s \in i} w_{ls} (1 + m_s)}{\sum_{l,s \in i} w_{ls}},
\end{equation}
where $m_s$ is the multiplicative bias factor for an individual source galaxy $s$ and $K$ is the weighted average of this multiplicative bias for the ensemble.

The second correction accounts for the shear responsivity, $\mathcal{R}$. This term corrects for the fact that galaxy ellipticities are a noisy proxy for the true gravitational shear. It is calculated for each radial bin $R_i$ as:
\begin{equation}
\mathcal{R} (R_i) = 1 - \frac{\sum_{l,s \in i} w_{ls} e_{\text{rms, s}}^2}{\sum_{l,s \in i} w_{ls}}.
\end{equation}
Typically, a value of $\mathcal{R} \approx 0.84$ is used to account for this in HSC WL analyses~\citep{medezinski2018planck}.


The statistical precision of these measurements is set by the covariance of the binned shear profile, which generally receives three contributions~\citep{gruen2015, umetsu2016}:
\begin{equation}
    \mathsf{C} = \mathsf{C}^{\mathrm{shape}} + \mathsf{C}^{\mathrm{lss}} + \mathsf{C}^{\mathrm{int}} .
\end{equation}
Here, $\mathsf{C}^{\text{shape}}$ represents the diagonal covariance due to statistical shape noise. The variance in bin $i$ is given by~\citep{miyaoka2018multiwavelength}
\begin{equation}
    \sigma^2_{\text{shape}} (R_i) = \frac{1}{ 4 \mathcal{R}^2 (R_i)}\frac{1}{[1 + K(R_i)]^2 \sum_i w_i}.
    \label{eq:shape_noise}
\end{equation}

The dominant term in this expression is the total weight of source galaxies in the bin, $\sum_i w_i$. This sum is roughly proportional to the number of galaxies in the bin, $N_{\text{gal}}(R_i)$. For logarithmic radial bins, the area of each annulus $A_i$ scales with radius. Assuming a uniform distribution of background sources, the number of galaxies per bin scales with the area of that bin, $N_{\text{gal}}(R_i) \propto A_i \propto R_i^2$. The shape noise variance in \cref{eq:shape_noise} is therefore $\sigma^2_{\text{shape}} \propto 1/N_{\text{gal}} \propto 1/R_i^2$. This means the uncertainty (standard deviation) has a radial dependence of:
\begin{equation}
\sigma_{\text{shape}}(R_i) \propto \frac{1}{\sqrt{N_{\text{gal}}(R_i)}} \propto \frac{1}{R_i}.
\label{eq:one_over_r}
\end{equation}
This $1/R$ trend is a key feature of the data: the innermost bins have the largest uncertainties, while the outermost bins are the most precisely measured.

The additional terms, $\mathsf{C}^{\text{lss}}$ and $\mathsf{C}^{\text{int}}$, represent the 
contribution from uncorrelated large-scale structure along the line of sight \citep{hoekstra2003well}, and the intrinsic variations in the lensing signal at fixed halo mass, arising from scatter in concentration, halo triaxiality, and correlated structures \citep{gruen2015}. Their relative importance depends on the scales probed. As found in \citet{miyatake2019weak}, the total uncertainty per cluster is dominated by the shape noise at $R \lesssim 3 \, \Mpch$, beyond which the contribution from $\mathsf{C}^{\text{lss}}$ becomes more important. The contribution from $\mathsf{C}^{\text{int}}$ matters most in the cluster core ~\citep{miyatake2019weak} but remains subdominant at all radii for our weak-lensing measurements. Since our analysis focuses on the regime where the error is shape-noise dominated, as shown in \citet{umetsu_cluster-galaxy_2020}, we restrict the covariance treatment to the $\mathsf{C}^{\text{shape}}$ component. This approximation captures the leading contribution to the statistical uncertainty without significantly biasing the results. We further discuss the validity of this assumption in \cref{sec:systematics_clusters}.

\begin{figure*}
\includegraphics[width=\textwidth]{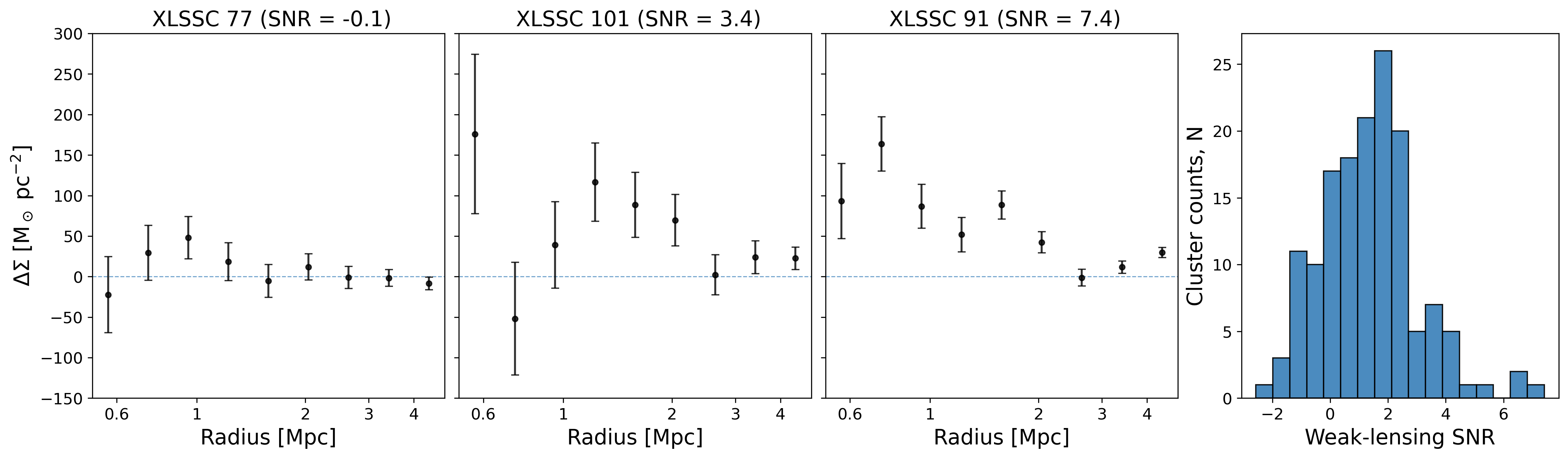}
\caption{Example weak-lensing ESD measurements for three clusters spanning the range of SNRs in the XXL–HSC sample. The first three panels show clusters with low (Cluster XLSSC~$77$, SNR $= -0.1$), medium (Cluster XLSSC~$101$, SNR $= 3.4$), and high (Cluster XLSSC~$91$, SNR $= 7.4$) weak-lensing SNRs. Points denote the measured ESD ($\Delta \Sigma$) with associated shape-noise uncertainties and the blue dashed marks $\Delta\Sigma = 0$. The fourth panel shows the distribution of weak-lensing SNR for all clusters in the sample. The median SNR is $1.25$.}
\label{fig:SNR}
\end{figure*}

We show the signal-to-noise ratio (SNR) distribution for the HSC-XXL cluster sample in \cref{fig:SNR}, as well as examples of the measured ESD for a high-, medium- and low-SNR cluster. The SNR is defined as SNR $= \langle d \rangle / \sigma_{\langle d \rangle}$ with
\begin{align}
    \langle d \rangle &= \frac{\sum_{i =1}^N \Delta \Sigma_{+} (R_i)/ \sigma_{\text{shape}}^2 (R_i)}{\sum{i =1}^N 1/\sigma^2_{\text{shape}}(R_i)},\\
\sigma_{\langle d \rangle} &= \frac{1}{\sqrt{\sum_{i =1}^N 1/\sigma^2_{\text{shape}}(R_i)}}.    
\end{align}
We follow \citet{umetsu2016} in choosing this linear definition over the commonly used quadratic signal-to-noise ratio $\text{SNR}_q \equiv [\sum (\Delta \Sigma_{i} / \sigma_{\text{shape},i})^2]^{1/2}$, as the quadratic form is always positive and therefore tends to yield spuriously high values when the measurement is noise-dominated. Our chosen linear estimator is a weighted average and can therefore take negative or positive signs. It retains the sign of the measured ESD, so ${\rm SNR}<0$ flags objects with an overall negative $\Delta\Sigma$.



\section{Exhaustive Symbolic Regression} \label{section:ESR}

Symbolic Regression (SR) is a machine-learning approach designed to find mathematical expressions to fit a given dataset. Unlike conventional regression, which relies on optimising parameters for a pre-defined expression, SR 
tries to uncover both the structure of the model and its parameters directly from the data. The aim is to find functions that accurately fit the data, but are also easy to interpret and generalise, giving some insight into the physical processes that generated the data \citep{Koza1992, SchmidtLipson2009}; see \citet{Kronberger_2024} for a recent review.

To do this, SR begins by selecting a set of mathematical operators (e.g. $x$, $+$, $-$, $\times$, $\sin$, $\exp$, $\log$) and combines them to generate candidate expressions. We define the complexity of an equation as the number of operators, parameters and variables it contains. Each candidate is then evaluated against the data and ranked according to a likelihood or loss function.

In recent years, a variety of SR algorithms have been developed. The most common approach is based on genetic programming \citep{Turing1950, goldberg1994genetic, cranmer_interpretable_2023}, in which simpler expressions that perform well are mutated or combined to create more complex candidates. These methods attempt to optimise the search process by avoiding trying sub-optimal solutions. Other approaches to SR include physics-motivated algorithms that exploit symmetries in the data \citep{udrescu2020ai, rene2021approach, keren2023computational}, as well as methods that incorporate neural networks to guide the search for optimal equations \citep{petersen2019deep, kim2020integration, valipour2021symbolicgpt}.

However, these machine learning methods are not infallible~\citep{la2021contemporary, bartlett_exhaustive_2024,Kronberger_2024_inefficiency}. Since they do not explore the entire space of possible functions, but rather focus on subsets deemed optimal, there is always a risk that the best solution may be missed. This uncertainty makes it difficult to assess the robustness of results obtained through SR. 


To address this, a new algorithm called Exhaustive Symbolic Regression (ESR) was developed\footnote{\url{https://github.com/deaglanbartlett/esr}} \citep{bartlett_exhaustive_2024}. Given a basic set of operators, ESR systematically considers every possible combination of operators that generate functions of a given complexity. This brute-force method is currently the only algorithm capable of guaranteeing that the optimal solution is found (within a given maximum complexity), since it considers all possible options within the allowed space. As such, it is particularly useful for solving simple problems where the best models are of low complexity \citep{desmond_functional_2023,sousa_optimal_2024}. At higher complexity, as the number of equations increases, ESR becomes more computationally expensive, but it can still serve as a good starting point for other stochastic methods.

ESR is described in detail in \citet{bartlett_exhaustive_2024}; here, we outline its main stages. First, it generates candidate models and optimises their free parameters according to a likelihood function. Second, it ranks these models using an information-theory-based metric called the Minimum Description Length (MDL).

\subsection{Generating functions}

To generate mathematical expressions, ESR begins by defining a basic set of operators. We consider three different types of operators: binary (e.g. $+$, $\times$), unary (e.g. $\exp$, $\log$) and nullary (parameters and variables). By representing these operators as nodes, functions can be visualised as trees~\citep{petersen2019deep}. The complexity of a function, previously defined as the number of operators and parameters, corresponds also to the number of nodes required to express an equation. 

Given a chosen level of complexity, ESR generates all possible trees with that number of nodes. It then populates the trees with all combinations of operators from the operator set, ensuring that connections follow the rules for binary, unary, and nullary operators. In ESR, the choice of complexity and the operator set are the only degrees of freedom.
In this project, we use the following operator set: 
\begin{equation}
    \{x, a, \text{inv}, \exp, \log, +, \times, -, /, \text{power}\},
\end{equation}
where $\text{inv}(x) \equiv 1 / x$, and $\text{power}(a,b)\equiv a^b$.

Once all possible trees are generated, the next step is to simplify the expressions and remove duplicates. Many expressions can be mathematically equivalent (for example, $\log(ax) = \log(a) + \log(x)$ for $a, x>0$), so eliminating duplicates ensures that each candidate function is unique and avoids redundant computation.
We then optimise parameters of the surviving candidate functions to find their maximum-likelihood values on the dataset in question. The method to achieve this is described in more detail in Section~\ref{optimisation}.

The number of possible functions grows rapidly with increasing complexity, so in practice there is a limit to how complex the expressions can be. In our case, we consider functions up to and including complexity $10$, corresponding to a total of $134239$ unique candidate expressions.

\subsection{Ranking functions}

Once a list of all possible mathematical models has been generated and fitted to the data, we require an appropriate metric to rank functions based on the quality of the fit. The simplest option is to rank functions according to the likelihood they give the data, $\mathcal{L}$, which quantifies how accurately they reproduce the data.

However, using just the likelihood is prone to overfitting: more complex functions tend to describe the data better but are more difficult to extrapolate to other datasets. Plotting each function on a complexity-likelihood plane, the functions that achieve the highest likelihood at each complexity  form what is known as the \textit{Pareto front}. These functions are referred to as \textit{Pareto optimal}. The Pareto front defines a set of optimal trade-offs but it does not provide a principled way of choosing one function over another within that set. To obtain a one-dimensional ordering of the models, we need a metric that maximises the likelihood while penalising complex hypotheses. 

In ESR, this is done by using the \textit{Minimum Description Length} principle ~\citep[MDL,][]{rissanen1978modeling, grunwald2007minimum, grunwald2019minimum}. MDL states that the best mathematical description of a dataset is the one that compresses the information the most, i.e. the function requiring the fewest units of information to encode both the model and the data. This is represented by the total \textit{Description Length}, $L(D)~=~L(D|H)+L(H)$, which consists of two terms:

\begin{enumerate}
\item $L(D|H)$ measures how accurately the hypothesis $H$ fits the data $D$. It describes the residuals of the data around the function’s expectation. This is related to the likelihood $\mathcal{L}$ via $L(D|H) = - \log \mathcal{L}$ under a Shannon–Fano      coding scheme \citep{cover1999elements}. 

\item $L(H)$ accounts for the complexity of the hypothesis $H$, penalising models with more operators and free parameters, especially those that have to be specified to high precision.
\end{enumerate}

Together, these terms ensure that overly simple functions (with low $L(H)$) are penalised for poor accuracy, while overly complex functions (with potentially low $L(D|H)$) are discouraged to prevent overfitting.

The full expression for $L(D)$ relevant for SR is derived in \citet{bartlett_exhaustive_2024} and is given by
\begin{equation}
    \begin{split}
        L(D) &= L(D|H) + L(H) \\
            &= - \log\mathcal{L}(\hat{\mathbf{\theta}}) + k\log(n) + p \log(2) + \sum_i^p \log(|\hat{\theta}_i|/ \Delta_i) \\ 
            &  \quad + \sum_j \log(c_j),
    \end{split}
    \label{eq:DL}
\end{equation}
where $k$ is the number of nodes to represent the function, $n$ is the number of unique operators, $p$ is the number of free parameters, 
$1 / \Delta_i$ is the precision with which parameter $i$ is specified and $c_j$ are constant natural numbers appearing in the function simplification process. The hat notation indicates evaluation at the maximum likelihood point and we denote the $p$ free parameters of the function as $\{\theta_i\}$. \cref{eq:DL} can be interpreted as an approximation to the Bayesian evidence under a certain choice of parameter prior and function prior
\citep{bartlett2023priors}. 


In the MDL metric, the balance between accuracy and simplicity depends on the strength of the data. With more data and greater precision, the likelihood term becomes more important than the complexity term. This means that with stronger data, the need for precision outweighs the penalty for complexity.

\subsection{Priors on functions}

The functions generated by ESR are purely mathematical expressions, without any inherent physical motivation. As a result, many of these functions exhibit structures that we do not often see in physical laws. For example, an expression such as $x^{x^{x}}$ is fairly simple according to \cref{eq:DL} and appears in the set of equations generated by ESR, but is unlikely to be relevant in most physics scenarios.

To guide ESR toward physically meaningful expressions, we can introduce a prior that encodes domain-specific knowledge about the types of equations typically encountered in physics. We achieve this by training a language model on a dataset of equations, allowing it to learn patterns and up-weight functions containing combinations of operators that are common in the dataset and hence more likely to be theoretically plausible.

In particular, we use a Katz back-off model\footnote{We use the implementation designed for SR given at \url{https://github.com/deaglanbartlett/katz}.} \citep{katz2003estimation, bartlett2023priors}. This is a probabilistic method originally developed in natural language processing to estimate the likelihood of word sequences. In our context, we instead use it to assign probabilities to sequences of mathematical operators based on their occurrence in a training corpus of equations.

To do this, the Katz model breaks down each candidate equation into $n$-grams: sequences of $n$ consecutive operators. It then examines how often each $n$-gram appears in the training set of equations. If a sequence appears frequently in the set, it is assigned a high probability. However, when an operator sequence is not found, the model ``backs off'' to a shorter $n$-gram until a match is achieved. This approach ensures that even expressions not explicitly present in the training set receive a non-zero probability, based on the frequency of occurrence of their components in the training set. For full details on the method see \citet{bartlett2023priors}.

By incorporating the Katz back-off model into ESR, we modify the MDL metric. Without the back-off model, the MDL contains the structural complexity term $k\log n$, which can be interpreted as a prior on the function that favours functions with fewer operators and distinct types of operator. This is combined with (an approximation to) the Bayesian evidence to achieve the total probability of the function (the negative logarithm of which is given by \cref{eq:DL}). When the Katz model is included, this prior is replaced with $- \log \Pi$, where $\Pi$ is the probability assigned to the function by the Katz model. This corresponds to assigning a shorter codelength to functions that are more common in the training corpus. In this way, the model favours equations that not only fit the data well but are also structurally consistent with 
functions that have found utility in physics.

Here, we study $n$-grams up to $n=10$, corresponding to the highest complexity considered for the functions. The corpus of equations we used for training the Katz model is taken from \citet{bartlett2023priors}. It includes the equations in the Feynman Symbolic Regression Database \citep{udrescu2020ai}, which were selected from the Feynman Lectures on Physics \citep{feynman1963feynman}, and a selection of $41$ equations taken from pages linked to Wikipedia’s ``List of scientific equations named after people'' \citep{Guimera_2020}.

\section{Methods}
\label{section:Methods}

\subsection{Modeling the ESD from 3D Mass Profiles}

To interpret the measured ESD in terms of the underlying matter distribution, we must connect $\Delta \Sigma(R)$ to the three-dimensional mass density, $\rho(r)$.

This can be done by projecting the density profile, $\rho(r)$, along the line of sight, $z$, to obtain the two-dimensional surface mass density, $\Sigma(R)$:
\begin{equation}
    \Sigma (R) = \int_{-\infty}^{\infty}{\rho \left(\sqrt{R^2 + z^2} \right) {\rm d}z},
\end{equation}
where $R$ is the projected radius, $z$ is the line-of-sight coordinate and $r^2 = R^2 + z^2$. Here, we assume that the mass distribution is spherically symmetric. The ESD can then be computed from its definition in \cref{eq:ESD_definition}.
For practical calculations, it is convenient to use a computationally simpler expression that avoids nested integrals~\citep{cromer_towards_2022}:
\begin{equation}
    \Delta \Sigma (R) = \frac{4}{R^2} \int_0^R {\rm d}r\, r^2 \rho(r)
    - 4R \int_0^{\pi/2} {\rm d}\theta\, \frac{\rho(R \sec \theta)}{4 \sin \theta + 3 - \cos(2\theta)} .
    \label{eq:rho_to_esd}
\end{equation}

In principle, the total excess surface density $\Delta \Sigma (R)$, had contributions from different components: the one-halo dark matter term, baryonic components, and the two-halo term. However, within the radial range probed in this analysis ($0.3 \leq R \leq 3.0 \, h^{-1} \mathrm{Mpc}$), the signal is dominated by the one-halo term. The two-halo term becomes significant only at scales spanning several virial radii, extending beyond our outer limit, while the baryonic contribution is restricted to the innermost regions. We discuss these components and the validity of these approximations further in \cref{sec:extensions}. Therefore, we define $\rho(r)$ as the density profile of the one-halo term.

\subsection{Function Optimisation and Parameter Estimation} \label{optimisation}

A key step in ESR is to optimise the free parameters of each candidate expression. ESR defines the best-fit parameters as those that maximise the likelihood of the data. We use an uncorrelated Gaussian likelihood of the form
\begin{equation}
    - \log \mathcal{L} = \sum_i \frac{\left(\Delta \Sigma^{\text{model}}(R_i) - \Delta \Sigma^{\text{obs}}_i \right)^2}{2 \sigma_i^2} ,
\end{equation}
where $\Delta \Sigma ^{\text{model}}(R_i)$ is the model-predicted ESD profile derived from the ESR expressions using \cref{eq:rho_to_esd}. $\{\Delta \Sigma^{\text{obs}}_i\}$ are the set of observed ESD at projected radii $\{R_i\}$ measured using the estimator in~\cref{eq:ESD}. Finally, the statistical uncertainties $\{\sigma_i\}$ are the errors from our ESD measurement. We use $\sigma_i = \sigma_{\text{shape}}(R_i)$ as defined in~\cref{eq:shape_noise}.

Two conditions are forced when optimising functions for them to be considered a viable density model:

\begin{enumerate}
    \item Positive density: the density function must be positive within the domain of the data, ensuring that the enclosed mass increases with radius. 
    \item Exclusion of profiles with constant terms: Any function containing constant terms in the density, such as $\rho(r) = \theta_0$, is excluded. Constant terms have no impact on the ESD (as can be seen from \cref{eq:rho_to_esd}).
    Including such profiles introduces redundancy, as they provide the same ESD as functions without the constant term but are more complex. Therefore, we do not consider them.


\end{enumerate}
These conditions are imposed during optimisation, with parameter values that violate them being assigned a likelihood of 0 and hence being discarded.

The numerical optimisation for local parameters is performed using a combination of two algorithms. The first is the \textsc{BFGS} algorithm \citep{Broyden1970,Fletcher1970,Goldfarb1970,Shanno1970,NocedalWright2006}, which is a gradient-descent method. The gradients are computed using automatic differentiation with \textsc{JAX} \citep{jax2018github}. While \textsc{BFGS} is fast and effective for well-behaved functions with a clear global minimum, it sometimes gets stuck in local minima when the objective function is more complex. To mitigate this problem, we repeat the optimisation $N_{\text{iter}}$ times with different random starting points and select the best result. We select random starting values for each $\theta_i$ from a uniform distribution within the range $[-10, 10]$. A solution is considered reliable if it converges to the same likelihood value at least $N_{\text{conv}}$ times.

If, after running the algorithm for $N_{\text{iter}}$ times, it fails to converge to a solution $N_{\text{conv}}$ times, we run a second minimisation routine: the \textsc{Nelder-Mead} algorithm \citep{nelder1965simplex}, which is a derivative-free, simplex search method. We find that this method is less likely to get lost in local minima as it makes larger jumps in parameter space. It does, however, take significantly longer to converge, which is the reason we do not use it from the outset. The Nelder-Mead algorithm is also run $N_{\text{iter}}$ times and considered reliable if it converges to the same solution in $N_{\text{conv}}$ of them.

After both optimisation stages, we examine the resulting parameter values. If any parameter exceeds $10^5$
 or falls below $10^{-5}$, we repeat the Nelder-Mead optimisation in log-space to enable a more effective search of regions with very large or small parameter values.

We found this routine to be the most effective at locating global minima, even for functions that are less well-behaved. For the examples shown here, we set $N_{\text{iter}} = 800$ for \textsc{BFGS} and  $N_{\text{iter}} = 700$ for \textsc{Nelder-Mead}. We choose $N_{\text{conv}} = 20p + 80$, where $p$ is the number of free parameters to be optimised. This scaling takes into account the increased difficulty of finding the global optimum in higher-dimensional spaces. 


Finally, to reduce the computational cost, we used a two-stage filtering strategy for complexities greater than $6$. First, we evaluated the full library of functions on the subset of $10$ clusters with the highest SNRs. We then selected the top $1000$ performing functions from this subset and optimised them across the full sample of clusters.

To ensure that we were not missing any important functions, we took the next $300$ functions (i.e., those ranked $1001$ to $1300$ for the $10$ clusters) and performed the full-sample optimisation on this set. We confirmed that none of these models appeared in the final top $100$ ranking for the full sample. Given that our results focus on the highest-ranked models (see Section \ref{section:results}), this test confirms that the filtering strategy is robust and that the excluded functions are indeed suboptimal and need not be considered.


\subsection{Uncertainty estimation}

The final step in computing the description length $L(D)$ is to determine the parameter precision, given by $1 / \Delta$. This is the only degree of freedom in the expression of the MDL. Increasing $\Delta_i$ (i.e., specifying $\theta_i$ with lower precision) reduces the parameter codelength term, $ \log(|\theta_i|/ \Delta_i)$, but at the same time increases the log-likelihood term by deviating from the best-fit parameter value.

In ESR, the $\Delta_i$ values for each parameter are estimated from the Hessian matrix of the log-likelihood. As shown in \citet{bartlett_exhaustive_2024}, the form of $\Delta_i$ that minimises the overall description length is $\Delta_i = \left(12 / I_{ii}\right)^{1/2}$, where $I_{ij} \equiv -\partial_i\partial_j\log\mathcal{L}\vert_{\hat{\theta}}$. This approach, which relies on the diagonal elements of the observed Fisher matrix to determine parameter precision, assumes that the likelihood function is approximately Gaussian near its maximum.
However, our model imposes constraints, such as enforcing a positive density profile, which can introduce sharp boundaries in the likelihood surface. In such cases, this Laplace approximation may not provide a reliable estimate of $\Delta_i$.

\begin{figure}
\includegraphics[width=80mm]{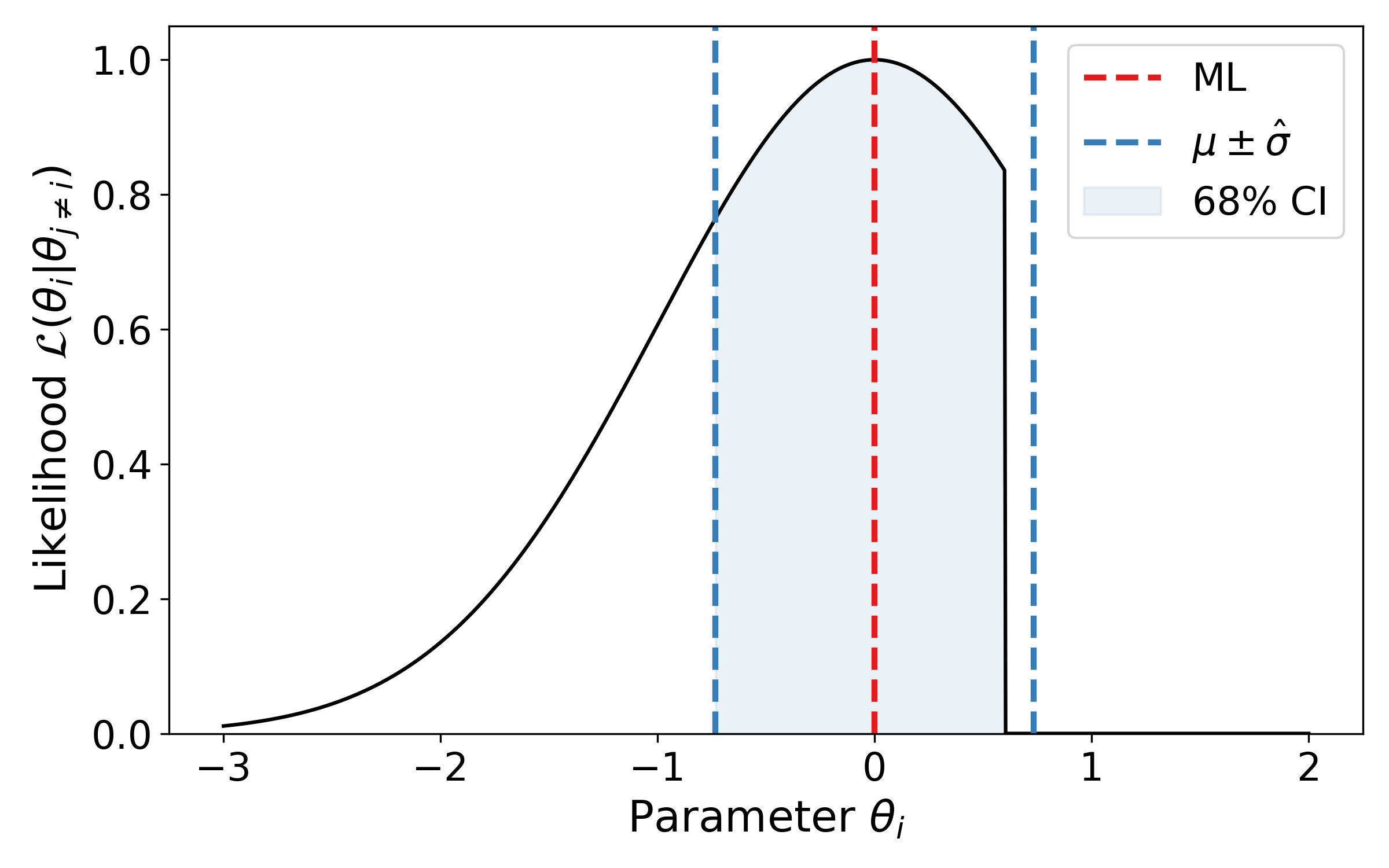}
\caption{Illustration of parameter uncertainty estimation from the conditional likelihood $\mathcal{L}(\theta_i)$. The red dashed line marks the maximum-likelihood (ML) estimate, and the shaded region corresponds to the 68\% confidence interval. For our analysis, we define a symmetric uncertainty $\pm\hat{\sigma}$ (blue dashed lines), where $\hat{\sigma}$ is set to the larger of the two distances from the ML value to the boundaries of the 68\% confidence interval.}
\label{fig:cutoff_likelihood}
\end{figure}

To handle these cases, we determine $\Delta_i$ directly from the likelihood surface. 
We fix all parameters except $\theta_i$ to their maximum likelihood (ML) values and compute the conditional (profile) likelihood $\mathcal{L}(\theta_i | \theta_{j \ne i} = \hat{\theta}_j)$. To estimate the uncertainty of this profile, we integrate the likelihood function for $\theta_i$ outward from the ML point in both directions, until the cumulative probability reaches $68\%$. If the likelihood has a hard cutoff on one side, we treat that cutoff as one of the integration limits. Since ESR requires symmetric uncertainty, we take the largest of the two one-sided as the standard deviation $\sigma_i$, and define $\Delta_i = \sqrt{12} \cdot \sigma_i$. \cref{fig:cutoff_likelihood} illustrates this procedure, showing how $\sigma_i$ is obtained by integrating outward from the ML value and then selecting the largest of the two resulting intervals.

In some cases, it might happen that $|\theta_i|/ \Delta_i<1$, meaning that the ML estimate of a parameter is indistinguishable from $0$ within the uncertainty tolerance. In these situations, we set these parameters to $0$, recalculate the likelihood, and reduce the parameter count $p$ by one. Since this occurs only when a parameter is already poorly constrained, the overall effect on $L(D)$ is minimal. Furthermore, if setting a parameter to $0$ causes the entire density profile to be $0$, we also set all remaining parameters to $0$ and reduce the number of parameters to $0$.

\subsection{Handling Multiple Datasets: Global vs. Local Parameters}

So far, we have described how ESR can be used to identify the best-fitting expression for a single dataset. However, since we are interested in studying dark matter halo profiles across a range of galaxy clusters, we want to be able to combine different datasets. This requires finding a systematic way to combine the results from individual clusters to evaluate overall model performance.

There are different ways to do this, depending on how the model parameters are treated. One approach is to fit the free parameters separately for each cluster and then combine the results. This involves applying ESR independently to each dataset and obtaining a separate fit of the density profile for each cluster. To compute a combined $L(D)$ across the full sample, we sum the likelihood contributions from each cluster while counting complexity terms only once. This yields a total $L(D)$ given by
\begin{equation} 
    \begin{split}
        L(D) &= - \sum_{a=1}^N  \log \mathcal{L}^{(a)} + k\log(n) + N \cdot p \log(2) \\ 
        & + \sum_{a=1}^N \sum_{i=1}^p \log(|\theta_i^{(a)}| / \Delta_i^{(a)}) + \sum_j \log(c_j).
    \end{split}
    \label{eq:DL_global} 
\end{equation}
In this case, since all clusters have the same functional form, we only need a single copy of the ``structural complexity'' terms (e.g. $k\log n$), and thus this is not simply the sum of the individual description lengths of each cluster.

Alternatively, we can consider the possibility that a subset of parameters is shared across all galaxy clusters; we refer to these as \textit{global} parameters. These act as universal constants across the full sample, representing the shared features in the shape of the dark matter density profile. An example of a parameter that could be treated as global is the inner slope of the density profile in the generalised NFW form. This approach of allowing for shared global parameters has been developed in other symbolic regression analyses, for example~\citealt{Tenachi2024ClassSR, Russeil2024MvSR, Fong2024MSR, Russeil2025ExploringMvSR}.

Introducing global parameters results in a lower total likelihood, since there are fewer degrees of freedom to fit each dataset. However, it also reduces the number of free parameters and thus the overall model complexity, which may make such models preferable under the MDL framework. Ideally, one would like to allow some parameters to be global while others remain galaxy-specific in an attempt to find the overall lowest description length models. This allows the data itself to determine which model structure is appropriate. The description length then becomes
\begin{equation} 
    \begin{split}
        L(D) &= - \sum_{a=1}^N  \log \mathcal{L}^{(a)} + k\log(n) + (N \cdot p_{\ell} + p_{\rm g}) \log(2)  \\ 
        & +\sum_{a=1}^N \sum_{i=1}^{p_{\ell}} \log(|\theta_i^{(a)}| / \Delta_i^{(a)})
        + \sum_{k=1}^{p_{\rm g}} \log(|\theta_k^*| / \Delta_k^*) \\
        & + \sum_j \log(c_j),
    \end{split}
\end{equation}
where $p_{\ell}$ and $p_{\rm g}$ are the number of local and global parameters, respectively, (with $p = p_{\rm \ell} + p_{\rm g}$) and the $*$ indicates global parameters.

Testing all possible combinations of global and local parameters is computationally very expensive. Instead, we use a two-step procedure to identify which combinations are worth trying. First, we fit each cluster independently, allowing parameters to vary between clusters. This ensures the best possible likelihood for each function. Then, we evaluate whether ``globalising'' some of the parameters can reduce the $L(D)$ enough to make the function a preferred model. We estimate the maximum possible gain in $L(D)$ under the best-case assumption that the likelihood remains unchanged upon globalisation. This upper bound is given by

\begin{equation}
    \begin{split}
        \text{max} (\Delta L) &= \text{max} (L_{\rm \ell} - L_{\rm g}) \\
        &=
        \sum_{a=1}^N \sum_{i=1}^{p_{\rm g}} \log \left(|\theta_i^{(a)}| / \Delta_i^{(a)}\right) +  p_{\rm g} (N - 1) \log 2,
    \end{split}
\label{eq:globalise}
\end{equation}
where the maximum gain corresponds to the minium possible contribution of the term $\sum_{k=1}^{p_{\rm g}} \log(|\theta_k^*| / \Delta_k^*)$. This minimum cost is achieved when a parameter's magnitude equals its resolution ($|\theta_i| = \Delta_i$). The ESR framework enforces this as a lower limit, as any parameter with a value smaller than its resolution ($|\theta_i| < \Delta_i$) is considered non-significant and is ``snapped'' to zero.
If the maximum gain is enough to situate the candidate expression in the top $10$ ranked functions we then perform the optimisation with global parameters. We choose this cut-off of rank 10 for practical reasons. First, this matches the number of functions that we report in our final analysis (see Table~\ref{tab:bestfitting}). Second, the description length $L(D)$ drops rapidly for lower-ranked functions, meaning they have negligible posterior probability ($\propto\exp(-L(D))$) and hence do not justify the extra computational cost of re-optimisation.

From a total search space of $134{,}239$ unique functions, our method identified $1411$ candidate expressions for globalisation. These candidates correspond to $867$ distinct functional forms with various combinations of global and local parameters.

\subsection{Optimising global parameters}

Unlike optimising purely local parameters, specific to each cluster, studying a combination of local and global parameters requires a different approach. One possible method is to perform a single optimisation over all $p_{\rm g} + N \cdot p_{\rm \ell}$ parameters, where $N=149$ is the number of clusters. However, in such a high-dimensional space, the optimiser often struggles to find the global optimum.

Instead, we use a two-step nested optimisation. The first optimisation searches for suitable global parameter values. For each trial set of global parameters, a secondary optimiser adjusts the local parameters for each cluster independently using $N_{\text{iter}} = 20$ and an $N_{\text{conv}} = 5$. This process repeats until the global optimiser converges $10$ times to the same global values. This method is more computationally expensive than a single optimisation, since local parameters are optimised even for suboptimal global values. However, it improves overall convergence by breaking the problem into smaller, lower-dimensional sub-problems.

For local parameter optimisation, we use the BFGS algorithm which is fast and generally effective in lower-dimensional settings. Global parameters are optimised using Nelder-Mead, which does not require gradient information and is more robust for exploring the global parameter space.


\section{Results} \label{section:results}

\subsection{Best-performing DM profiles}

We show in \cref{tab:bestfitting} the best-performing density profiles obtained using ESR across all galaxy clusters. The functions are ranked using \cref{eq:DL}. The description length is decomposed into three components: residuals between the data and the function, complexity of the functional form, and parameter contributions (as detailed in the table footnote). For comparison, we also include the ranking obtained using the Katz prior. In that case, the total description length is computed using the residual and parameter terms, with the structural complexity term replaced by the value from the Katz column.

We find that the function rankings are very similar under both criteria. Since we are using a dataset of $149$ galaxy clusters, the residual and parameter terms, both of which scale linearly with the number of clusters, dominate over the complexity term. With more data, the complexity penalty becomes less important, since overfitting is increasingly disfavoured by the data itself. As a consequence, adopting a Katz prior produces only minor differences in the final rankings. For this reason, in the rest of the paper we focus on the rankings obtained without the Katz prior.

We also report in \cref{tab:bestfitting} the relative probability $P(f_i \mid D)$ of each model with respect to all other candidates. In the MDL framework, the $L(D)$ of model $i$ (i.e. function $f_i$) is (up to an additive constant) the negative logarithm of its probability given the data, hence
$P(f_i\!\mid D)\propto \exp(-L(D)_i)$. We therefore convert the $L(D)$ for each function $f_i$ into a 
relative probability:
\begin{equation}
P(f_i | D) = \frac{\exp{\left(-L(D)_i\right)}}{\sum_j \exp{\left(-L(D)_j\right)}},
\label{eq:Prel}
\end{equation}
where the sum is over all functions we analysed up to complexity $10$. This shows that the relative probability is dominated by the first model, which carries $0.835$ of the weight.

\begin{table*}
    \centering
    \begin{tabular}{|c|c|c|c|c|c|c|c|c|c|c|}
        \hline
        \multirow{2}{*}{Rank} & \multirow{2}{*}{Rank Katz} & \multirow{2}{*}{$\rho(r) / 10^{12} M_{\odot} \text{Mpc}^{-3}$} & \multirow{2}{*}{Complexity} & \multirow{2}{*}{$P(f|D)$} & \multicolumn{6}{c|}{Description Length} \\
        \cline{6-11}
        & & & & & Residual$^1$ & Parameters$^2$ & Function$^3$ & Katz$^4$ & Total & \shortstack{Total \\ Katz} \\
        \hline
        1 & 1 & $1/(\theta_0(r|\theta_1|)^{r^2})$ & 10 & 0.84 & 601.4 & 223.5 & 16.1 & 16.8 & 841.0 & 841.6  \\
        2 & 5 & $1/(\theta_0(-r + |\theta_1|^{r^r}))$ & 10 & 0.16 & 620.0 & 204.8 & 17.9 & 25.1 & 842.6 & 849.8 \\
        3 & 3 & $(|\theta_0r|^r \cdot |\theta_1|)^{-r}$ & 10 & 0.01 & 601.3 & 229.1 & 16.1 & 17.3 & 846.5 & 847.7 \\
        4 & 20 & $1/(\theta_0(r^r|\theta_1|)^r)$ & 10 & $4.7 \times 10^{-5}$ & 622.3& 212.3 & 16.1 & 25.7 & 850.7 & 860.4 \\
        5 & 2 & $(1/(r^2|\theta_0|))^r/\theta_1$ & 10 & $4.2 \times 10^{-5}$ & 615.7 & 219.1 & 16.1 & 11.0 & 850.8 & 845.7 \\
        6 & 8 & $r^2/(\theta_0|\theta_1|^r)$ & 9 & $3.5 \times 10^{-5}$ & 616.5 & 222.1 & 12.5 & 15.5 & 851.0 & 854.0 \\
        7 & 7 & $(r^{1/r}|1/\theta_0|)^r/\theta_1$ & 10 & $2.0 \times 10^{-5}$ & 623.3 & 212.7 & 16.1 & 16.9 & 851.6 & 852.4 \\
        8 & 13 & $r/(|\theta_0|^r|\theta_1|)^r$ & 9 & $ 5.1 \times 10^{-6}$ & 622.2 & 216.3 & 14.5 & 17.4 & 853.0 & 855.9 \\
        9 & 4 & $(|\theta_0|^r|\theta_1|/r)^{-r}$ & 10 & $4.1 \times 10^{-6}$ & 605.2 & 231.9 & 16.1 & 12.2 & 853.2 & 849.2 \\
        10 & 16 & $r/(\theta_0|\theta_1|^r)$ & 7 & $3.9 \times 10^{-6}$ & 623.8 & 219.8 & 9.7 & 14.1 & 853.2 & 857.7 \\
        
        \vdots & \vdots & \vdots & \vdots & \vdots & \vdots & \vdots & \vdots & \vdots & \vdots & \vdots \\
        221 & 338  & NFW 1: $\theta_0 / (r (|\theta_1| + r)^2) $ & 9 & $\sim 10^{-28}$ & $685.3$ & $206.8$ & $24.0$ & $12.6$ & $904.7$ & $916.2$ \\
        903 & 895 & Isothermal: $\theta_0 /r ^2 $ & 5 & $\sim 10^{-56}$ & $944.5$ & $19.0$ & $5.5$ & $5.3$ & $968.9$ & $968.7$ \\
        2436 & 2361  & NFW 2: $\theta_0 / (r (1/|\theta_1| + r)^2) $ & 9 & $\sim 10^{-110}$ & $939.1$ & $126.9$ & $26.0$ & $12.6$ & $1092.0$ & $1078.6$ \\
        2964 & 2925 & gNFW: $\theta_0 / (r (|\theta_1| + r)^{\theta_2})$  & 9  & $\sim 10^{-159}$ & $793.8$ & $384.1$ & $26.9$ & $19.4$ & $1204.7$ & $1197.3$ \\
        3251 & 3251 & Einasto: $\theta_0 \exp{(\theta_1 r^{\theta_2})}$ & 8 & $\sim 10^{-209}$ & $953.2$ & $355.1$ & $12.9$ & $11.6$ & $1321.2$ & $1319.9$ \\
        3507 & 3509 & Burkert: $\theta_0 / ((\theta_1 + r)(\theta_1 + (r)^2) $ & 13 & $< 10^{-300}$ & $707.9$ & $1310.1$ & $31.2$ & $20.7$ & $2038.7$ & $2049.2$ \\
        3528 & 3527 & DZ: $\theta_0 / (r^{\theta_1}(\theta_2 + r^{1/2})^{2(3.5 - \theta_1)})$  & 17  & 0 & $850.8$ & $1275.0$ & $41.7$ & $32.5$ & $2167.53$ & $2158.3$ \\
        \hline
    \end{tabular}
    \begin{tabular}{c}
        $^1 - \sum_a \log\mathcal{L}^{(a)} ( \hat{\bm{\theta}} )$ \qquad\qquad $^2 p \log(2) + \sum_{a=1}^N \sum_{i=1}^p \log(|\theta_i^{(a)}| / \Delta_i^{(a)})$
        \qquad\qquad $^3 k \log n + \sum_j \log(c_j)$ \qquad\qquad $^4$ $- \log \Pi$  + $\sum_j \log(c_j)$ \\
    \end{tabular}
    \caption{Top functions found by ESR applied to the sample of $149$ HSC galaxy clusters, ranked by total description length. The total description length is decomposed in three terms: the accuracy term (``Residuals''), the parameters complexity term (``Parameter'') and the structural complexity term (``Function''). To calculate the total description length using the Katz prior (``Total Katz''), we replace the ``Function'' column with the ``Katz'' column. All functions presented here use exclusively local (i.e. separate per-cluster) parameters. For each model we also report the relative probability $P(f_i | D)$, obtained by normalising $\exp(-L(D)_i)$ over all functions. We additionally include several commonly used literature profiles: NFW \citep{navarro1997universal}, generalised NFW (gNFW; \citealt{evans}), Einasto \citep{einasto}, Burkert \citep{Burkert_1995}, and Dekel–Zhao \citep{DekelZhao}), written in the ESR functional language and evaluated using the same description-length formalism. 
    \label{tab:bestfitting}.
    }
\end{table*}

For reference, \cref{tab:bestfitting} also lists several common dark matter density profiles alongside their ranking. For complexities above $6$, we do not evaluate the full ESR function set on every cluster. Instead, we run all functions on the $10$ clusters with the highest signal-to-noise ratio, and then apply only a subset of the best-performing functions from this stage to the full cluster sample. This approach is sufficient to identify the best-fitting functions, but it may omit some lower-ranked candidates which would therefore not be included in this ranking.
This implies that the reference profiles might in fact rank worse if the full set were exhaustively evaluated. \cref{tab:bestfitting} already includes functions both with purely local parameters and with combinations of local and global parameters. The ranking reflects the full description length including accounting for loval versus global as described in \cref{section:Methods}.

The NFW profile is included in two different parametrisations. NFW $1$ uses a more standard form, written as $\rho(r) = \theta_0 / (r (|\theta_1| + r)^2)$. However, for $11$ clusters, the parameter precision for $\theta_1$ gives $\Delta_1 \geq \theta_1$, causing ESR to snap $\theta_1$ to $0$. This results in a profile of the form $\rho(r) \propto 1/ r^3$ which no longer reflects the behaviour of an NFW profile. To avoid this, NFW $2$ uses an alternative form $\rho(r) = \theta_0 / (r (1/|\theta_1| + r)^2)$. In this parametrisation, $\rho(r) \rightarrow 0$ as $\theta_1 \rightarrow 0$, which preserves the expected NFW behaviour when the scale radius becomes very small. Usually, when ESR finds two variants of the same function, it retains only the one with the better description length. However, because the NFW $1$ variant with a snapped parameter is not representative of the standard NFW profile, we have chosen to present both variants in our results. 

Among the commonly used profiles, the best-performing after NFW $1$ is the isothermal sphere. This is not due to a higher likelihood; rather, for $130$ out of the $149$ clusters, the model parameters are snapped to $0$, resulting in a smaller overall description length. The model effectively behaves as an isothermal sphere for the high-SNR clusters and as a null profile for the remainder.

The generalised NFW (gNFW) profile introduces an additional free parameter, $\theta_2$, which controls the outer slope of the density profile \citep{evans}. In principle, the extra flexibility should improve the likelihood relative to the standard NFW model. In practice, however, ESR often snaps poorly constrained parameters to zero for clusters with low signal-to-noise. After snapping, the resulting likelihood can in fact be \textit{worse} than that of the standard NFW model. Fitting a three-parameter profile locally to clusters that have only $9$ radial data points is therefore prone to overfitting and leads to parameters that are either weakly constrained or whose constraints likely reflect overfitting. We find a very broad and highly skewed distribution of inferred $\theta_2$ values, with many clusters favouring extreme and seemingly unphysical outer slopes. As a result, per-cluster estimates of $\theta_2$ should be interpreted cautiously. We have also included the Dekel-Zhao model \citep{DekelZhao} (DZ in \cref{tab:bestfitting}) and the Einasto \citep{einasto} and Burkert \citep{Burkert_1995} profiles, which have more flexible shapes but perform worse.

More importantly, the best-fitting functions found by ESR consistently outperform the standard dark matter profiles in both likelihood and total description length while using the same number of free parameters as NFW. Interestingly, no functions with more than $2$ parameters were ranked among the top solutions. This suggests that the constraining power of the current dataset is limited to constraining only two free parameters effectively. This also explains why standard profiles with more than two parameters are systematically less preferred by the MDL framework.

\begin{figure}
\includegraphics[width=90mm]{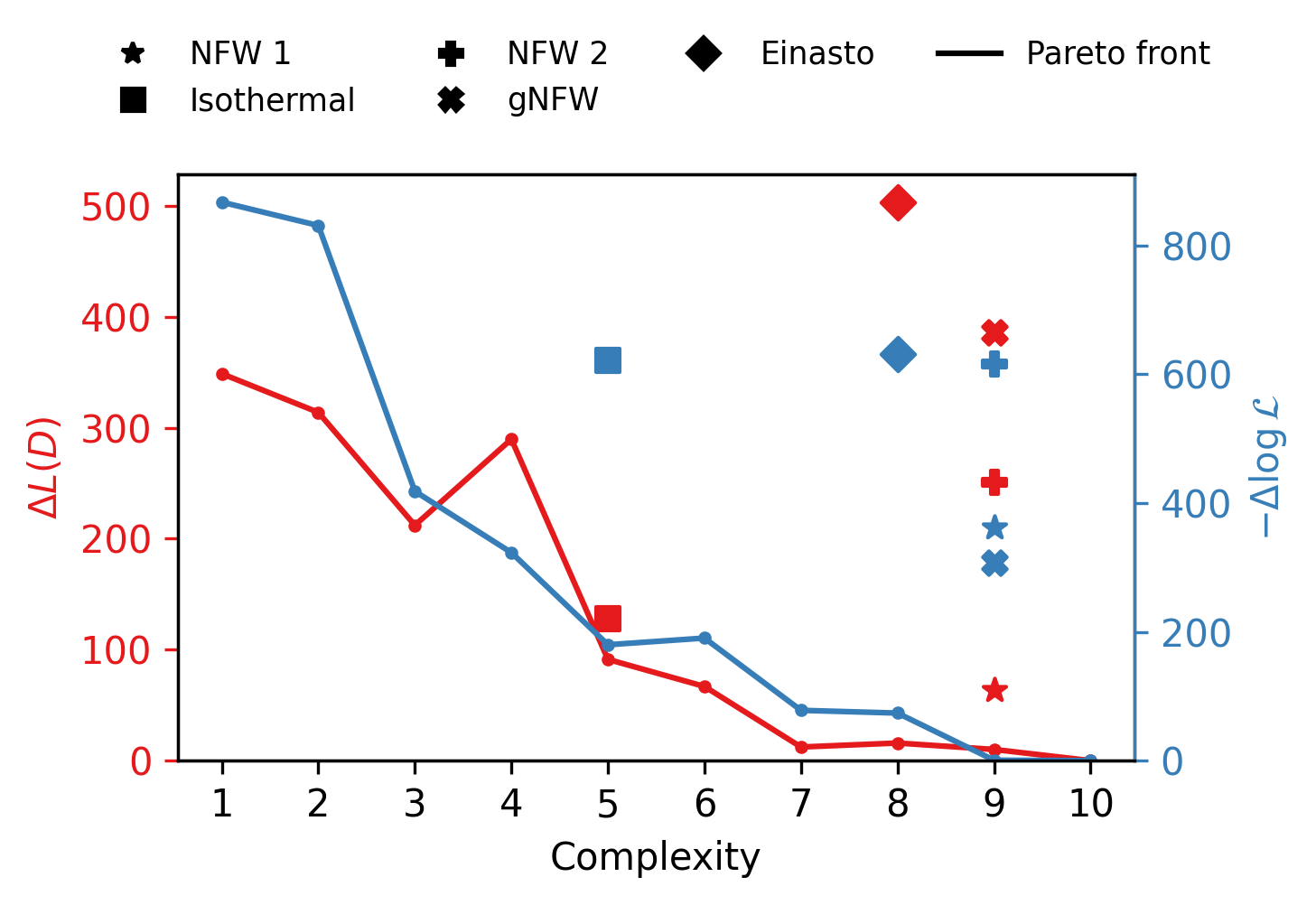}
\caption{Best-fitting functions at ecah complexity according to the change in description length, $\Delta L (D)$, and the likelihood, $\mathcal{L}$, relative to the corresponding minima. The markers show the position in the $L(D)$ plane (red) and likelihood plane (blue) of some common dark matter halo profiles. The Burkert and Dekel-Zhao profiles are not included here for clarity, since they have very poor $L(D)$. Only the $L(D)$ values without the Katz prior are shown, since including the Katz prior produces very similar results. The two NFW points correspond to two parametrisations used in the analysis. NFW $1$ is the standard form, 
$\rho(r)=\theta_0 / (r(|\theta_1|+r)^2)$ and NFW $2$ is $\rho(r) = \theta_0 / (r (1/|\theta_1| + r)^2)$.}
\label{fig:paretos}
\end{figure}

\cref{fig:paretos} shows the best
description length, $L(D)$, and the negative log-likelihood, $- \log(\mathcal{L})$, at each model complexity. For each metric, the results are normalised by subtracting the global minimum value, so that the globally best functions appear at $0$ and worse-performing functions appear higher up. The plot shows that the common dark matter density profiles are Pareto-dominated by the functions discovered by ESR. Note that beyond complexity $7$, the improvement in description length is small; however, complexity $10$ remains the best. Further improvement may be possible at complexities higher than $10$, which is currently beyond the scope of ESR. As expected, the negative log-likelihood continues to decrease with added complexity. For completeness, we note that the function with the single best raw likelihood is a $4$-parameter, complexity-$10$ model:
\begin{equation}
    \rho(r) = \left| \frac{\theta_1}{\theta_0 + r} + \theta_2 \right|^{\theta_3}.
\end{equation}
The behaviour of this expression is sensitive  to the signs and magnitudes of the parameters. The term $\theta_1/(\theta_0 + r)$ introduces a potential divergence when $\theta_0 + r \rightarrow 0$, which can occur at a finite radius if $\theta_0$ is negative. Conversely, a zero in the density occurs at the radius at which $\theta_1/(\theta_0 + r) + \theta_2 = 0$. Both features, a divergence and zero-density shells, appear for some clusters in the set depending on the fitted parameter values. These behaviours are allowed by the formula and such narrow spikes or zeros are difficult to detect numerically and avoid during fitting. Although this model attains the best raw likelihood among all expressions considered, it is not ranked as a top solution because it is penalised for its high number of free parameters. This shows how a highly flexible expressions that can overfit noisy data are naturally down-weighted by MDL.


All functions in \cref{tab:bestfitting} use local parameters. To the list of best-fitting functions with local parameters, we applied \cref{eq:globalise} to identify functions and parameters worth globalising. This produced a list of $2905$ functions and parameter combinations. \cref{tab:funcs_global_params} lists the best-performing functions that include some global parameters and their corresponding rankings. This table also includes the gNFW profile with the inner power slope treated as a global parameter. The best value of the global parameter $\theta_2$ found is $3.61 \pm 0.05$.  


Interestingly, the best-fitting functions with global parameters are generally simpler. This is because global parameters force the function to fit all clusters at once using the exact same values. While complex functions are usually more flexible, locking their parameters makes them too rigid to handle the differences between individual clusters. Simpler functions, by contrast, work better here because they capture the general average of the sample rather than struggling to fit the unique details of each cluster.

The top-ranking function from this set is found at rank $38$ in the full list of functions where any parameter can be global or local, indicating the globalising parameters does not lower the description lengths of good functions in this case.
This is perhaps unsurprising given the mathematical form of the best-fitting profiles. Global parameters are expected to be useful when they describe a feature of the profile itself rather than individual clusters, such as a simple scaling or powers of the radius. However, in the best-fitting functions, the parameters do not correspond to simple scaling relations or power laws, so cluster-specific parameters perform better.

\begin{table*}
    \centering
    \begin{tabular}{|c|c|c|c|c|c|c|c|c|c|}
        \hline
        \multirow{2}{*}{No.} & \multirow{2}{*}{Rank} & \multirow{2}{*}{$\rho(r) / 10^{12} M_{\odot} \text{Mpc}^{-3}$} & \multirow{2}{*}{Complexity} & \multicolumn{6}{c|}{Description Length} \\
        \cline{5-10}
        & &  & & Residual$^1$ & Parameters$^2$ & Function$^3$ & Katz & Total & Total Katz \\
        \hline
        1 & $38$ & $|\theta_0|^{-|\textcolor{red}{\boldsymbol{\theta_1}}|^r}$ & 6 & 753.98 & 105.06 & 14.48 & 5.68 & 873.53 & 879.21 \\
        2 & $49$ & $(r|\theta_0|)^{-1/|\textcolor{red}{\boldsymbol{\theta_1}}|^r}$ & 9 & 753.32 & 106.05 & 20.92 & 19.02 & 880.30 & 899.32 \\
        3 & $50$ & $(r|1/\theta_0|)^{|\textcolor{red}{\boldsymbol{\theta_1}}|^{-r}}$ & 8 & 756.02 & 104.80 & 19.71 & 19.67 & 880.53 & 900.20 \\
        4 & $53$ & $(r|1/\theta_0|)^{|\textcolor{red}{\boldsymbol{\theta_1}}|^r}$ & 7  & 759.10 & 109.16 & 14.48 & 11.67 & 882.75 & 894.42 \\
        5 & $54$ & $r^{\theta_0}/\textcolor{red}{\boldsymbol{\theta_1}}$ & 5 & 761.11 & 112.73 & 9.70 & 7.85 & 883.54 & 891.39 \\
        \vdots & \vdots & \vdots & \vdots & \vdots & \vdots & \vdots & \vdots & \vdots & \vdots \\
        -- & 2335  & gNFW $\theta_0 / (r (|\theta_1| + r)^{\textcolor{red}{\boldsymbol{\theta_2}}}) $ & 9 & $888.83$ & $158.47$ & $26.88$ & $19.42$ & $1074.18$ & $1066.72$ \\
        -- & 3289  & gNFW $1 / ( \theta_0 r (|\theta_1| + r)^{\textcolor{red}{\boldsymbol{\theta_2}}}) $ & 9 & $679.85$ & $661.56$ & $29.19$ & $20.74$ & $1370.60$ & $1362.15$\\ 
        \hline
    \end{tabular}
    \caption{Top-ranked functions with some global parameters. The information displayed is the same as in \cref{tab:bestfitting}. Parameters set to global are highlighted in bold red. For comparison, the gNFW with $\theta_2$ as a global parameter is also included.  The ``Rank'' column shows each function's position relative to the complete list of all models analysed in this work. \label{tab:funcs_global_params}}
\end{table*}

\subsection{Properties of the best-fitting ESR functions}



All of the functions in \cref{tab:bestfitting} provide a good fit to the data, but their physical interpretation is not necessarily clear. Two general features stand out when examining the shapes. First, most functions tend to $0$ as $r \rightarrow \infty$ as expected, although their inner behaviour can differ from one profile to the other. Second, most of these functions are not analytically integrable, making it impossible to obtain a closed-form expression for the enclosed mass.

Commonly used profiles such as Einasto do not provide a closed-form expression for the enclosed mass, so the lack of an analytical mass estimate is not in itself a drawback. However, for practical applications it is often useful to work with profiles that are both analytically integrable and reasonably simple while still having competitive fits. \cref{tab:integrable} lists some of these profiles. In general, these functions have lower complexity than the best-fitting ones and are easier to handle analytically, while still greatly outperforming  literature functions such as NFW.

\begin{table*}
    \centering
    \begin{tabular}{|c|c|c|c|c|c|c|c|c|c|}
        \hline
        \multirow{2}{*}{No.} & \multirow{2}{*}{Rank} & \multirow{2}{*}{$\rho(r) / 10^{12} M_{\odot} \text{Mpc}^{-3}$} & \multirow{2}{*}{Complexity} & \multicolumn{6}{c|}{Description Length} \\
        \cline{5-10}
        & & & & Residual$^1$ & Parameters$^2$ & Function$^3$ & Katz & Total & Total Katz \\
        \hline
        1 & 6 & $r^2/(\theta_0|\theta_1|^r)$ & 9 & 616.47 & 222.08 & 12.48 & 15.47 & 851.03 & 854.02 \\
        2 & 9 & $r/(\theta_0|\theta_1|^r)$ & 7 & 623.76 & 219.76 & 9.70 & 14.14 & 853.22 & 857.66 \\
        3 & 21 & $1/r + r/(\theta_0|\theta_1|^r)$ & 10 & 623.64 & 217.88 & 17.92 & 19.60 & 859.44 & 861.12 \\
        4 & 22 & $1/(\theta_0 \cdot(r + (\theta_1 - r)/r))$ & 10 & 399.38 & 442.62 & 17.92 & 15.54 & 859.92 & 857.54 \\
        5 & 32 & $(r^{-2} + r/\theta_0)/\theta_1$ & 10 & 641.83 & 214.02 & 16.09 & 10.33 & 871.94 & 866.18 \\
        \hline
    \end{tabular}
    \caption{This table presents a subset of functions that are analytically integrable and thus have analytical mass profiles. As in \cref{tab:funcs_global_params}, the ``Rank'' column shows each function's position relative to the complete list of all models analysed in this work. \label{tab:integrable}}
\end{table*}

We are going to highlight a few profiles from this list and illustrate their properties and potential drawbacks. The first profile we focus on is the highest-ranked function in \cref{tab:integrable} (ranked 6th overall):
\begin{equation}
\rho(r) = \frac{r^2}{\theta_0|\theta_1|^r} = \rho_0 r^2 \exp{(- \lambda r)},
\label{eq:best_integrable}
\end{equation}
The second expression provides a more intuitive re-parameterisation with a normalization $\rho_0 = 1/\theta_0$ and a decay constant $\lambda = \ln|\theta_1|$ (assuming $|\theta_1| \neq 0$). The profile rises as $r^2$ from the origin before being cut off by an exponential decay. Its enclosed mass profile is given by:

\begin{equation}
M(R) = \frac{96\pi\rho_0}{\lambda^5} \left[ 1 - e^{-\lambda R} \left( 1 + \lambda R + \frac{(\lambda R)^2}{2} + \frac{(\lambda R)^3}{6} + \frac{(\lambda R)^4}{24} \right) \right].
\end{equation}

The primary physical drawback of this function is the unphysical ``hole'' at its centre, where $\rho(0)=0$. We find that this feature is common among the analytically integrable profiles discovered by SR (see \cref{tab:integrable}). Therefore, we also highlight a second profile that solves this issue. This profile is ranked third in \cref{tab:integrable} (rank 21 overall) and has the form
\begin{equation}\rho(r) = \frac{1}{r} + \frac{r}{\theta_0|\theta_1|^r} = \frac{1}{r} + \frac{r}{r_\star} \exp \left( - \lambda r \right)
.\label{eq:second_integrable}
\end{equation}
 This profile diverges as $\rho(r) \sim 1/r$ for small $r$, similar to the inner cusp of an NFW profile. Here, the second form is a more physically intuitive re-parameterisation, where the decay constant is $\lambda = \ln|\theta_1|$ and the scale radius is $r_\star = \theta_0$. This profile's behaviour is well-aligned with standard halo models: it diverges as $\rho(r) \sim 1/r$ for small $r$, creating an inner cusp similar to an NFW profile, while the second term ensures a rapid exponential decay at large radii. 

A potential problem with this function derived by ESR is that it is not dimensionally consistent. We can approach this in two ways. The first option is to introduce a constant of proportionality, which transforms the expression into a physical three-parameter model:
\begin{equation}
\rho(r) = \frac{A}{r} + B r \exp \left( - \lambda r \right)
\label{eq:profile_cusp_physical}
\end{equation}
where $A$, $B$, and $\lambda$ are the three free parameters to fit. While this ensures that the equation is now physically valid, adding a third parameter increases the complexity term of the $L(D)$. Specifically, optimising \cref{eq:profile_cusp_physical} across all clusters yields a residual negative log-likelihood of $-\log \mathcal{L} \approx 573.32$. This represents an improvement over \cref{eq:second_integrable} due to the model's added flexibility. However, the complexity penalty for the extra parameter outweighs this gain, resulting in a total description length of $L(D) = 1740.63$. The alternative is to interpret the numerical coefficients in the ESR expression as having implicit dimensions that balance the equation. While this avoids adding new parameters, it effectively ``hard-codes'' the units of the data into the function constants, so the equation is no longer unit-agnostic.

The corresponding integrated mass profile $M(r)$ is given by:
\begin{equation}
M(r) = 2\pi A r^2 + \frac{24\pi B}{\lambda^4} 
- 4\pi B e^{-\lambda r} \left( \frac{r^3}{\lambda} + \frac{3r^2}{\lambda^2} + \frac{6r}{\lambda^3} + \frac{6}{\lambda^4} \right).
\label{eq:mass_profile_cusp}
\end{equation}
For the HSC dataset, these profiles offer a good compromise, providing a competitive fit while retaining the practical advantages of an analytical mass profile.

\begin{figure*}
\includegraphics[width=\textwidth]{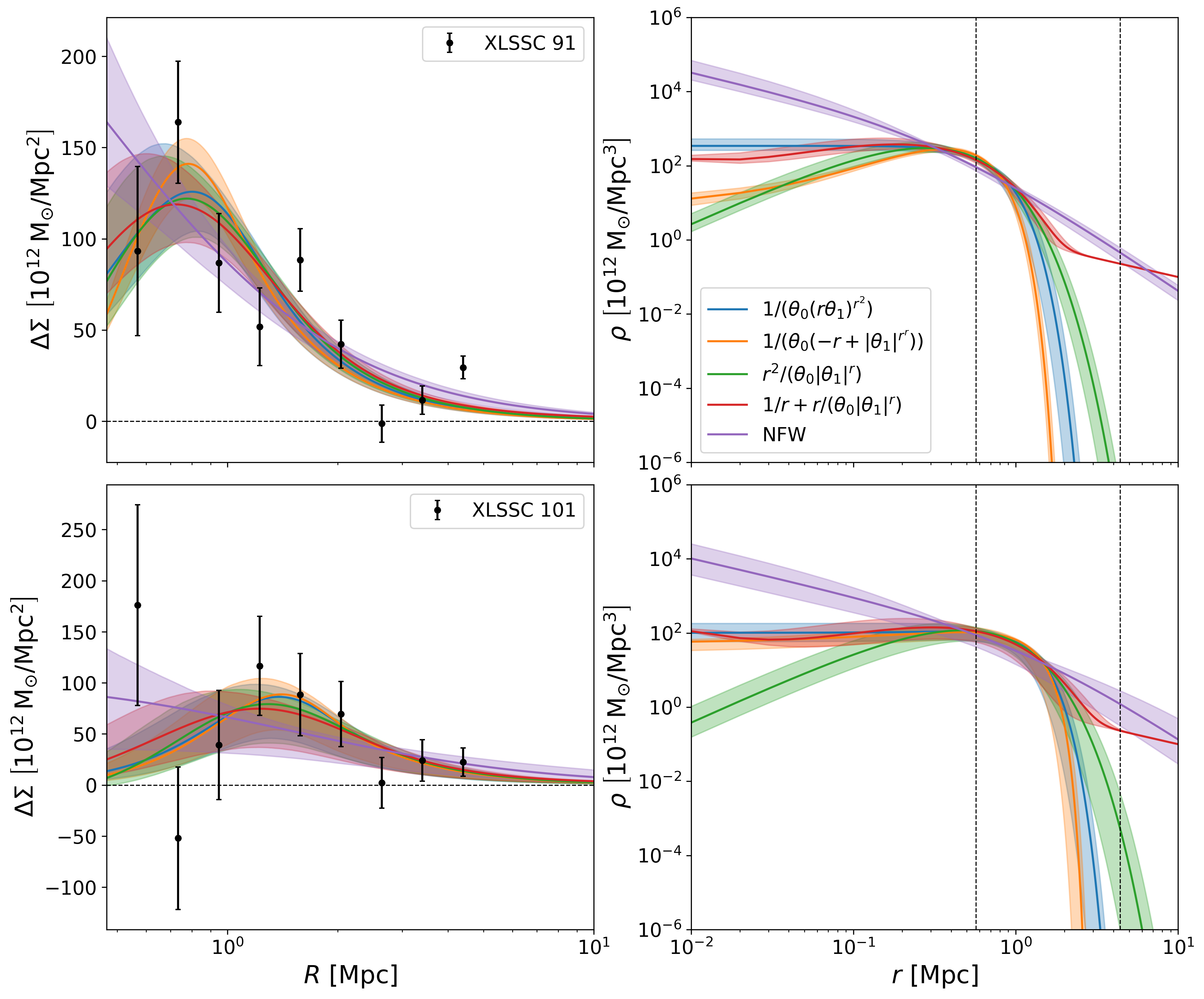}
\caption{Plots of the top $2$ best-fitting functions, together with the top analytically integrable functions discussed in the text. The NFW profile is also shown for comparison. The top panels correspond to cluster XLSSC~$91$, which has the highest SNR in the sample, while the bottom panels show cluster XLSSC~$101$, which is a medium SNR cluster. These two clusters are also shown in \cref{fig:SNR}.
\emph{Left panels:} ESD fits to the data. The horizontal dashed line marks zero.
\emph{Right panels:} Corresponding density profiles plotted over a wide radial range. The vertical dashed lines indicate the radial range covered by the data. Shaded regions represent the $68\%$ credible intervals derived from the posterior parameter distributions.
}
\label{fig:best_funcs}
\end{figure*}

\cref{fig:best_funcs} illustrates the shape of the best-fitting models: the top three from \cref{tab:bestfitting} and the analytically integrable profile in \cref{eq:best_integrable}. The cluster shown in the top panel is XLSSC~$91$, which has a relatively high SNR of $7.41$, while the bottom panel shows cluster XLSSC~$101$, with a lower SNR of $3.40$. This allows us to compare the behaviour of the models for both high- and medium-SNR clusters.

An interesting feature of the ESR fits is the presence of a local maximum in the ESD within the range of the data. Unlike traditional models that usually have a smooth, monotonic decline, the data seems to prefer functions that allow for this turning point. Physically, this implies central regions where the density is either flat (core-like behaviour) or even decreasing at small radii. The data seems to prefer these inner behaviours rather than steep central cusps. The flexibility of the models also allows this feature to be placed at different radii. As the turnover is placed further away form the clusters also develops a more extended core.

However, the radial position of this feature suggests that it is also influenced by how measurement uncertainties vary with projected radius, $R$. As shown in \cref{eq:one_over_r}, the shape noise of the signal scales roughly as $\sigma_{\Delta\Sigma} \propto 1/R$. This $1/R$ trend is clearly visible in the data shown in \cref{fig:best_funcs}: the error bars are largest in the innermost radial bins and become progressively tighter at larger $R$. Since the likelihood term weights each data point by its inverse variance ($1/\sigma_{\Delta\Sigma}^2$), the fit is more driven by the high-precision data points at large radii. As such, the data prefer functions that accurately capture the outer behaviour over constraining the inner shape. For example, for clusters with a high SNR like XLSSC $91$ (top panels in \cref{fig:best_funcs}) the maximum is placed close to the centre. For clusters with a lower SNR like XLSSC $101$ (bottom panels in \cref{fig:best_funcs}), the maximum shifts outward to better fit the data in the cluster outskirts.

This also explains the significant diversity of inner slopes among the top-ranked functions, as the current data lack the constraining power to distinguish between different profile functions at small radii.

The analytically integrable profile from \cref{eq:second_integrable} provides an interesting example of this behaviour. Mathematically, the $1/r$ term is expected to dominate the 3D density profile ($\rho$) at both very large and very small radii. Its influence at large radii ($r \gtrsim 1 \text{ Mpc}$) is evident in \cref{fig:best_funcs} (right panels). However, the $1/r$ divergence at small radii is not visible within most of the plotted range. For the best-fit parameters found by ESR, the second term, $r/(\theta_0 |\theta_1|^r)$, dominates the profile's shape throughout the intermediate region ($0.01 \lesssim r \lesssim 1 \text{ Mpc}$). This term is responsible for the ``bump'' and the subsequent turnover seen in the density plot. The $1/r$ inner cusp only begins to dominate at extremely small radii (approximately $r \lesssim 10^{-2} \text{ Mpc}$).

\subsection{Enclosed mass estimation}
\label{sec:enclosed_mass}

A practical application of the ESR-derived functions is to calculate the enclosed mass of the clusters. This is interesting because it avoids the need to rely on a single assumed density profile, as is often done with NFW. Instead, we take advantage of the set of best-fitting functions identified by ESR and their respective posterior probabilities to produce mass estimates that effectively marginalise over the uncertain functional form of the density profile.

To combine models, we use a weighting scheme based on the description length, $L(D)$. We convert the $L(D)$ for each function $f_i$ into a normalised probability weight, the relative probability, $P(f_i | D)$, (see \cref{eq:Prel}). This weighting ensures that models providing the most efficient description of the data contribute more, while those with poorer $L(D)$s are suppressed. We include functions in the average until the cumulative probability reaches $0.9999$, which corresponds to the first six entries in \cref{tab:bestfitting}.

For each of these top-ranking models, we quantify the parameter uncertainties by sampling the posterior distributions using a Hamiltonian Monte Carlo (HMC) algorithm \citep{Neal2011}. This is implemented with the \textsc{NumPyro} package \citep{phan2019composable}, which utilises the efficient No-U-Turn Sampler (NUTS) \citep{hoffman2014no}. We adopt uniform priors in logarithmic parameter space. Specifically, each model parameter is sampled as $\log_{10}|a_i| \sim \mathcal{U}(\log_{10}\hat a_i - 5 \sigma_i,\ \log_{10}\hat a_i + 5 \sigma_i)$, where $\hat a_i$ and $\sigma_i$ are the ESR optimiser best-fit and its associated uncertainty, respectively.
We run two independent chains with $2000$ warm-up steps and $6000$ samples. We confirm that the chains have converged by calculating the Gelman-Rubin statistic ($\hat{R}$) for all parameters and requiring it to be less than $1.01$ \citep{gelman1992inference}.

From each posterior sample we compute the corresponding enclosed mass profile. We define the enclosed mass of a cluster as the mass within a radius $r_{200}$, where the mean enclosed density is equal to $200$ times the critical density $\rho_{\text{crit}}(z)$ at the cluster redshift $z$. The enclosed mass at this radius is defined as $M_{200}$.  We determine the value of $r_{200}$ for each sample by numerically solving this condition and then evaluate $M_{200}$ at that radius. However, for certain functional forms or specific parameter combinations sampled by the HMC algorithm, the mean density never reaches the $200\rho_{\text{crit}}$ threshold (for instance, if the profile flattens). Since $r_{200}$ is mathematically undefined in these cases, these samples are considered unphysical and are discarded.

This procedure yields, for each function and each cluster, a posterior distribution of $M_{200}$. Finally, we combine the distributions across all functions by weighting each posterior mass estimate according to the $P(f_i | D)$ of its parent function. Specifically, if function $f_i$ has a model probability $P(f_i | D)$ and $n_i$ retained HMC samples, then each mass draw for $f_i$ is assigned a weight $P(f_i| D)/n_i$.
The resulting weighted distribution provides our final posterior for $M_{200}$, from which we report the mean and the $16$th/$84$th percentiles as uncertainty bounds.

\begin{figure}
\includegraphics[width=\columnwidth]{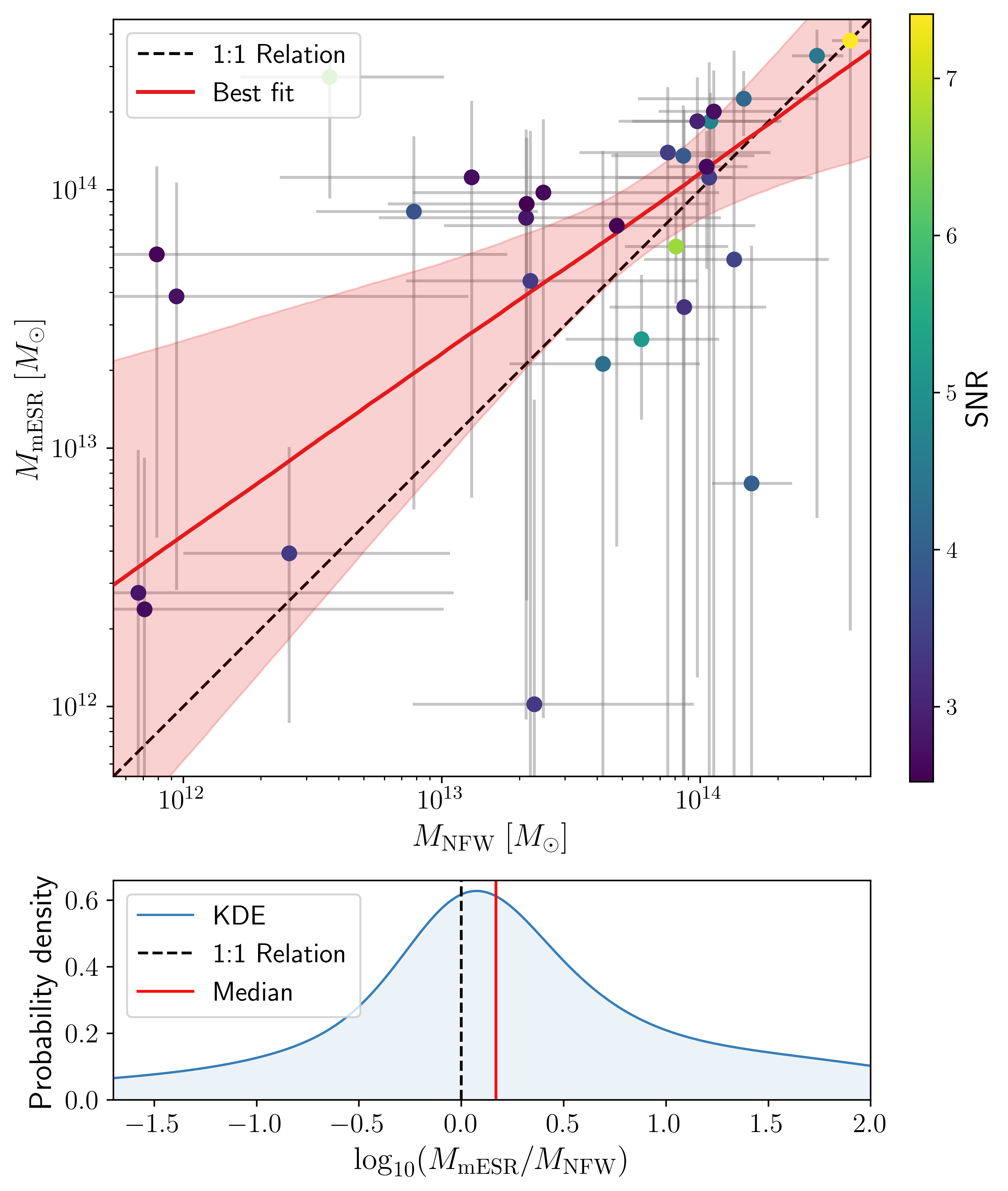}
\caption{Comparison of enclosed mass estimates, $M_{\mathrm{mESR}}$, obtained from the weighted combination of our best-fitting functions against those derived from the standard NFW profile, $M_{\mathrm{NFW}}$ for XXL clusters with a signal-to-noise ratio (SNR) higher than $2.5$. The top panel shows the results in log--log space. The dashed black line indicates the 1:1 relation, while the solid red line shows the best-fit linear relation obtained using the \textsc{roxy} package, with the shaded region representing the $1\sigma$ confidence interval.
The bottom panel shows the stacked posterior distribution of the logarithmic mass ration $\delta_i = \log_{10}(M_{\mathrm{mESR}}/M_{\mathrm{NFW}})$. Each cluster contributes its individual posterior, built by propagating the uncorrelated uncertainties on both mass estimates under a Gaussian approximation. The black dashed line shows the $1:1$ relation and the solid red line marks median of the stacked distribution, $\langle\delta\rangle = 0.17$\,dex.}
\label{fig:masses}
\end{figure}

For comparison, we repeat the same procedure using the NFW profile with the parametrisation NFW~1, given by $\theta_0 / (r (|\theta_1| + r)^2)$. The top panel in \cref{fig:masses} shows the weighted mean values of enclosed mass obtained from our best-fitting functions against the corresponding NFW estimates. 
We restrict the plot to clusters with an $\text{SNR} > 2.5$. These are a subset of $30$ clusters that provide sufficiently reliable signals. In the low-SNR regime ($\text{SNR} \le 2.5$), the data lack statistical power to meaningfully constrain mass models. This is evident from the standard NFW fits; for instance, for $11$ of these low-SNR clusters, the NFW fit collapses to a trivial $\rho = 0$ solution. In cases where such zero-snapping does not occur, fits to this low-SNR data result in extremely broad or unconstrained posterior distributions for $M_{200}$. We therefore focus only on this high-SNR subset to plot the correlation between the two mass estimates.

In the figure, the 1:1 relation is shown as a dashed black line. We also fit a linear relation between our marginalised ESR mass ($M_{\text{mESR}}$) and the NFW mass ($M_{\text{NFW}}$) in base-10 logarithmic space (dex). The model is given by:
\begin{equation}
\log_{10} M_{\text{mESR}} = \alpha + \beta \log_{10} M_{\text{NFW}}.
\end{equation}
We include a Gaussian intrinsic scatter $\sigma_\text{int}$.
We perform this regression using the \texttt{Python} package \textsc{roxy}\footnote{\url{https://github.com/deaglanbartlett/roxy}} \citep{bartlett2023marginalised}, which implements the "Marginalised Normal Regression" (MNR) algorithm. This method is specifically chosen because, as demonstrated by \citet{bartlett2023marginalised}, it provides unbiased results by robustly accounting for measurement uncertainties in both the $x$ and $y$ variables, the intrinsic scatter and the unknown distribution of the ``true'' $x$-values.

For the inputs, \textsc{roxy} requires symmetric errors. We therefore define the uncertainty in the logarithm of each mass estimate as half the difference between its $84$th and $16$th percentiles. We place improper uniform priors on the intercept $\alpha$ and slope $\beta$, and an improper uniform prior on $\sigma_\text{int}$. The latent true $x$-values are modelled with a Gaussian hyperprior, with its mean and width also inferred using improper flat hyperpriors. We ran the sampler with $700$ warm-up steps followed by $5000$ production samples, which yielded effective sample sizes $n_\text{eff} > 2900$ and Gelman-Rubin statistics $\hat{R} = 1.00$ for all inferred parameters.

From this fit, we recover a slope of $\beta = 0.75 \pm 0.50$, an intercept of $\alpha = 3.53 \pm 6.87$, and an intrinsic scatter of $0.29 \pm 0.17$ dex. While the slope is shallower than unity, it remains statistically consistent with a $1:1$ relation given the large uncertainty. Combined with the consistency of the intercept with 0, we conclude that both mass estimates yield consistent results.
Performing the same regression on the full sample of all clusters yields very similar results with a slope again consistent with $1$.


This consistency demonstrates that our model-averaging technique produces physically reasonable mass estimates comparable to the standard NFW approach. The bottom panel of \cref{fig:masses} shows the distribution of the logarithmic mass ratios, $\delta_i = \log_{10}(M_{\text{mESR}}/M_{\text{NFW}})$, which quantifies the relative deviations between the two masses. This shows the stacked posterior on this ratio over all clusters in the high-SNR sample, where the width of each posterior is derived by propagating the uncertainty on $\delta_i$ from the (uncorrelated) uncertainties on both mass estimates under a Gaussian approximation.
The final distribution is centred near zero (indicated by the dashed black line), confirming that our method does not introduce a significant systematic bias compared to the standard NFW model. The median of the stacked distribution is $\langle\delta\rangle = 0.17^{+0.93}_{-0.69}$\,dex (solid red line),
indicating that masses derived by ESR are on average higher than the NFW masses. 

We note that the MDL function has a relative probability of $0.835$ (see \cref{fig:best_funcs}). Consequently, this single profile is the dominant driver of the fit. To asses the influence of the MDL function in the mass inference, we compute the ratio $\delta_{MDL, i} = \log_{10}(M_{\text{mESR}}/M_{\text{MDL}})$, where $M_{\text{MDL}}$ denotes the enclosed masses inferred using the highest-ranked MDL model. The stacked distribution then has a median of $\langle\delta\rangle = 0.06$\,dex, showing that MDL-only masses are very similar to the model-averaged ESR masses.

We can also compare the precision of NFW masses to the weighted-ESR masses. To do this, we examine the relative uncertainties. On average, the fractional uncertainties of ESR-derived masses are $9 \%$ larger than those of NFW. This increase has two contributions: the statistical uncertainty within each model and the additional variance from averaging over different profile shapes. To separate these effects, we compare the uncertainties of the NFW masses directly to those obtained from the MDL model alone. We find that MDL masses are $1\%$ larger than those derived from NFW, implying that the dominant source of the increased ESR uncertainty comes from averaging between different models. In other words, NFW yields slightly tighter model-specific constraints, but the broader uncertainties in the ESR masses reflect the additional model uncertainty accounted for by the profile averaging.


We note, however, that the mass estimates for some of the clusters are lower than the values obtained by \citealt{umetsu_weak_2020}. The difference is a consequence of the different modelling assumptions. In \citealt{umetsu_weak_2020}, informative log-uniform priors are imposed directly on the halo mass and concentration, which favour physically reasonable solutions even for clusters with data consistent with zero. In contrast, our approach does not impose priors on physical parameters such as halo mass or concentration. Instead, the optimisation is driven by the likelihood, and the subsequent posterior sampling explores the region around the best-fit value. Consequently, our method is free to explore regions of the parameter space that may be disfavoured by physical priors in other works, if that is where the global minimum of the data lies.


on the mass and concentration, our method fits all model parameters freely. This means that the optimiser finds the values that best fit the data, without being constrained to regions expected by the cosmological model. While this can give different mass estimates than prior-informed methods, it provides a more direct measure of the information contained in the data itself. More importantly, the same methodology was applied to both the NFW and ESR-weighted masses, ensuring that the comparison shown in \cref{fig:masses} is self-consistent and presents a fair comparison of the models' performance on the data. 

This also highlights a more general methodological point. Relying solely on a single profile, such as NFW, carries a risk: if that chosen model is not a perfect description of the true halo structure, the resulting mass estimates may be systematically biased and reflect the priors (both functional and parametric) rather than the information contained in the data itself. We have shown here that not imposing any functional priors yields masses that are higher on average. The flexibility of our multi-model approach, with models motivated by the minimum description length principle, is that it can mitigate this model-selection bias by averaging over a set of plausible density profiles. In the Bayesian context this is called Bayesian model averaging.

\subsection{Evidence for a universal profile}

Another interesting question we can address using ESR and the MDL principle is whether the data favour a \textit{universal} density profile across all clusters, or whether a better description is obtained by allowing each cluster to have its own profile. Up to this point, our analysis has enforced a single profile shared by all clusters, while only varying global and local parameters. Here, we test what happens if we allow more than one functional form to describe the density profile of dark matter haloes.



To account for the possibility that different clusters are described by different profiles we modify the total description length to include the information cost of assigning a specific function to each cluster. The total description length becomes:

\begin{eqnarray}
L(D) = \sum_{a=1}^N L_{a,m(a)} + \sum_{m\in\mathcal{M}} \Phi_m + L_{\mathrm{assign}},
\end{eqnarray}
where $m(a) \in \{1,…,M\}$ denotes the profile assigned to cluster $a$, and $\mathcal{M}$ is the set of $M$ distinct profiles actually used in the sample.
The per-cluster term $L_{a,m}$ is the residuals plus parameters contribution evaluated for cluster $a$ under profile $m$, such that
\begin{equation}
\begin{split}
        L(D)_{a,m} &=  - \log\mathcal{L}(\mathbf{\hat{\theta}_m}) + p_{m} \log(2) + \sum_i^{p_m} \log(|\hat{\theta}_{m,i}|/ \Delta^{(a)}_{m,i}).
    \end{split}
\label{eq:Lam}
\end{equation}
This is the same as the usual \cref{eq:DL} just excluding the structural (function) penalty.
The structural penalty is then counted once per distinct profile with the term $\sum \Phi_m$, where $\Phi_m = k \cdot \log n + \sum_j \log(c_j) $ as before (see \cref{eq:DL}) and the sum is calculated over the $K$ unique profiles used across the sample.

Finally, $L_{\mathrm{assign}}$ encodes the cost of communicating which cluster uses which profile, encoding the ``label'' for each cluster. To minimise this cost, we use a coding scheme that assigns shorter binary labels to profiles that appear frequently in the population and longer labels to rare ones \citep{Shannon1948}.
Formally, let $n_m$ denote the number of clusters assigned to profile $m$ (where $\sum_m n_m = N$). If we assign a binary label of length $l_m$ to profile $m$, the total cost of transmitting the assignment vector for all $N$ clusters is 

\begin{eqnarray}
    L_{\mathrm{assign}} = \sum_{m = 1}^K n_m l_m.
\end{eqnarray}
To ensure that a continuous string of these binary labels can be parsed into a unique sequence of profiles, we use a \textit{prefix-free} code \citep{MacKay2003}. This requirement imposes Kraft’s inequality on the code lengths, given by $\sum e^{-l_m} \leq 1$ \citep{CoverThomas2006}. Minimising the total cost $L_{\mathrm{assign}}$ subject to this constraint gives the optimal Shannon code lengths,

\begin{eqnarray}
    l_m = - \log(n_m / N),
\end{eqnarray}
which essentially corresponds to using shorter codes for more frequent labels \citep{Shannon1948}. This gives a total assignment cost of
\begin{eqnarray}
L(D)_{\text{assign}} = - \sum_m n_m \log(n_m / N).
\end{eqnarray}

This is zero if all clusters share one profile and increases as more profiles are added. It penalises models that require a more complex or less uniform assignment of profiles across clusters. With this scheme, we try the two alternatives:

\subsubsection{Per-cluster profile}

First, we test the ``per-cluster'' assignment, where each of the $N=149$ clusters is free to select its own functional form. Because each cluster contains only $9$ radial data points, minimising the full description length on an individual basis tends to favour trivial solutions (i.e., $\rho=0$). To avoid this, we instead select the function that minimises the sum of the residuals and parameter costs for each cluster. This ensures that the fits remain penalised for unnecessary parameters, but prevents the selection from being driven entirely by the functional complexity term.

Once each cluster has selected its preferred function, we calculate the total description length, including the costs of encoding the functional forms and the assignment vector. This approach yields a total description length of $L(D) = 1752.27$, which is significantly worse than that of the best universal profile ($L(D) = 840.95$), as shown in \cref{tab:universal_comparison}.

Statistically, this model fails because the assignment cost ($L_{\text{assign}}$) is extremely high ($518.71$ nats). Given total freedom, the sample ``fragments'' into $M=51$ unique profiles. While the fit quality (the $\sum L(D)_{am}$ term) is better than for the universal profile -- as expected from the fact that we are choosing functions that minimise this quantity -- the functional cost from specifying different functions and the assignment term result is a worse total description length.

Physically, this result is consistent with the expectation that dark matter haloes are largely self-similar. Enforcing a common profile increases the statistical power to constrain the functional form, which far outweighs the flexibility gained by fitting clusters independently.

\subsubsection{Two-profile mixture model}

A natural compromise between the universal model and the unrestricted per-cluster assignment is to limit the flexibility to a small number of distinct profiles. In this scenario, we allow the sample to be described by a mixture of just two functions, where each cluster selects the best option from a pair of candidates.

In practice, we test all possible pairs formed from the $20$ best-performing functions listed in \cref{tab:bestfitting}. For each pair, every cluster independently selects the function that minimises the sum of the residual and parameter costs. We then compute the total description length of the entire sample, including the structural costs of the two functions and the assignment entropy.

The optimal pair identified by this search consists of the $2$nd and $3$rd highest-ranked profiles from the universal analysis:
\begin{eqnarray}
    \rho_{2}(r) = 1/(\theta_0(-r + |\theta_1|^{r^r})), \quad \rho_{3}(r) = (|\theta_0r|^r \cdot |\theta_1|)^{-r}.
    \nonumber
\end{eqnarray}
Because these already dominate the ranking (and because the next-best pairs correspond to combinations of the top few functions) we do not expect lower-ranked functions to outperform this combination.

It may seem counter-intuitive that the top-ranked universal profile is absent from this optimal pair, as it represents the best single-model compromise between accuracy and simplicity. However, inspection of \cref{tab:bestfitting}, shows that the functions at ranks $2$ and $3$ offer complementary advantages. The profile $\rho_2$ has lowest parameter description length ($204.8$) in the top set. Conversely, $\rho_3$ has the lowest residuals ($601.3$) and thus the highest likelihood. By partitioning the sample, the mixture model of these two profiles provides a better description length than including the rank 1 function.

For the optimal pair, the total description length is  $L(D) = 861.71$. This is a massive improvement over the per-cluster model ($L(D) = 1752.27$), but it is still worse than the single universal profile ($L(D) = 840.95$).

As expected, the $\sum L(D)_{am}$ term gives a better value than the best universal profile. However, this gain is outweighed by the penalties associated with the cost from having to specify two functional forms and encoding their assignments across the sample. In this case, $\rho_2$ is selected by $62$ clusters and $\rho_3$ by $87$ clusters, giving an assignment cost of $L_{\text{assign}} = 101.17$. Therefore, the single universal profile remains the simplest and statistically preferred description of the entire sample.

This analysis could be extended in different ways. First, one could explore mixtures of three or more profiles ($M \ge 3$). Second, the optimisation strategy itself could be refined. Currently, we use this approach where each cluster selects the profile that minimises its local fit (residuals + parameters). A more rigorous approach would be to perform a global optimisation, searching for the specific assignment of clusters to profiles that minimises the \emph{total} mixed-model description length directly. However, given the dominance of the universal profile and the rapidly increasing penalties for assignment and structure, these more complex models are unlikely to outperform the single-profile solution.

\begin{table*}
\centering
\begin{tabular}{l c c c c}
\hline
\hline
Model Type & Residuals + Parameters & Function Cost & Assignment Cost & \textbf{Total $L(D)$} \\
 & $\sum L(D)_{am}$ & $\sum \Phi_m$ & $L_{\text{assign}}$ & \\
\hline
Universal ($M=1$) & 824.85 & 16.09 & 0.00 & \textbf{840.95} \\
Two profiles ($M=2$) & 769.05 & 34.01 & 101.17 & 904.23 \\
Per-Cluster ($M=51$) & 633.70 & 538.08 & 518.71 & 1752.27 \\
\hline
\end{tabular}
\caption{Comparison of total description length for different profile–assignment models. Here $M$ 
denotes the number of distinct density profiles allowed in the model: $M=1$ corresponds to a single universal profile shared by all clusters, $M=2$ allows clusters to choose between two profiles, and $M=51$ is the per–cluster model in which each cluster is free to select its own functional form. The table reports the contributions from the residuals + parameter terms, the structural (function) costs for all unique profiles used, and the assignment cost required to encode which clusters use which profile. All values are given in nats. The total description length is shown in the final column.}
\label{tab:universal_comparison}
\end{table*}

\section{Discussion} \label{section:discussion}

\subsection{Interpretation of the ESR results}

In this work, we have presented an empirical derivation of dark matter density profiles using ESR. One of our main findings is that when the data are allowed to choose from a broad range of profiles, NFW is not among the best-fitting models. We have found other two-parameter models (such as the analytically integrable profiles in \cref{eq:best_integrable} and \cref{eq:second_integrable}) that the data prefers. This shows that functions with only two free parameters are sufficient to describe the current data, while also suggesting that HSC-XXL weak-lensing data favour density profiles that deviate systematically from the standard NFW shape. Although data with a SNR or extended radial coverage might prefer more complex functional forms, two parameters are optimal by the MDL criterion for this weak lensing dataset.


This raises the question whether the deviation from the NFW form is a property of the data or an artefact of the method. We addressed this in \citealt{martín2025constraining}, where we validated the method using synthetic clusters generated from an NFW profile. These mock clusters were modelled to closely resemble the HSC-XXL sample used here in radial sampling and mass distribution. Modelling the noise as a fractional uncertainty across the data, we tested different noise conditions and data quantities. Since ESR exhaustively explores the functional space up to the specified complexity, the NFW profile is always evaluated at complexity $9$. The question, therefore, is whether the MDL metric correctly identifies it as the optimal function. We found that ESR reliably identifies NFW among the top ten functions even when the ESD measurements have fractional uncertainties as large as $80\%$ (for searches up to complexity $9$). This shows that the generating function is identifiable under the MDL framework even in the presence of substantial statistical noise, provided that the dominant systematics are properly accounted for.

In contrast, when applying the methodology to the real cluster sample (and extending the search to complexity $10$), the NFW profile is not found among the top-performing functions. Even if we restrict the analysis to complexity $9$ and remove all complexity $10$ functions, the NFW profile ranks only $129$th, with a relative probability (\cref{eq:Prel}) of $P(f_{NFW}|D) \sim 10^{-24}$.
Consequently, the fact that NFW is highly ranked in controlled mocks but disfavoured in the real data strongly suggests that the true density profiles of these clusters do deviate from the NFW form. 

That said, the form of the preferred profiles and their physical origin must be interpreted with care. One possible interpretation is that the underlying dark-matter distribution may still resemble an NFW-like profile, but baryonic physics modifies the total mass density in ways that shift the preferred functional form away from pure NFW. For example, energetic AGN feedback can evacuate gas from the central regions of clusters, producing a depletion feature in the total mass profile \citep{delpopolo2019correlationsmatterdistributionclash, martizzi, Arjona_G_lvez_2024}. Such processes could potentially generate structures similar to the central holes or turnovers seen in some of the best-fitting ESR models. 

However, it is important to take into account the intrinsic sensitivity of the data. The weak lensing data has a limited constraining power at small radii compared to cluster outskirts.
Weak-lensing uncertainties decrease roughly as $1/R$ due to the radial scaling of the source-galaxy shape noise. As a consequence, ESR naturally favours functions that perform well in the outer regions of clusters, where the data have smaller uncertainties. Given this, while baryonic modification could play a role, the lensing data alone cannot unambiguously attribute the observed functional differences to any specific physical process. We discuss this further in \cref{sec:extensions}.


A further strength of our approach is its transparency in the low-signal-to-noise regime. Standard single-profile fits can return apparently well-defined mass estimates even when the ESD measurements are statistically consistent with zero signal, because the posteriors are shaped largely by the assumed profile. In contrast, ESR correctly identifies such cases as unconstrained and consistent with a null density.

\subsection{Broader implications}


Galaxy clusters are among the most powerful cosmological probes as their abundance as a function of mass and redshift is sensitive to the amplitude of matter fluctuations ($\sigma_8$) and the matter density ($\Omega_{\rm m}$) \citep{allen_2011}. The constraining power of cluster abundance measurements, however, is limited by the accuracy of mass calibration \citep{vikhlinin_2009, rozo_2010}. Because cluster masses are not observed directly, survey analyses rely on observable proxies, such as X-ray luminosity, optical richness, or the Sunyaev--Zel’dovich (SZ) signal \citep{pratt_2009, planck_2016_clusters}. To map these measurements to an underlying mass measurements, scaling relations between the survey observable and cluster mass must be carefully calibrated \citep{giodini_2013, von_der_Linden_2014}.



Weak gravitational lensing is commonly used to calibrate this mass proxies, as it measures the total gravitating matter and is independent of the dynamical state of the cluster gas. Nevertheless, weak lensing analyses still rely on modelling assumptions, like the choice of a parametric density profile, which can introduce systematic biases if the assumed form does not adequately describe the true mass distribution.

The potential biases introduced by profile mis-specifications have previously been investigated by simulations. For example, \citealt{Becker_2011} and \citealt{Linden_giants} demonstrated that fitting an NFW profile to more realistic, triaxial, or substructured halos typically results in a negative mass bias of approximately $5-10$ per cent.

Here, we have shown how this systematic might also arise in observational data. We computed $M_{200}$ using the best-fitting ESR functions and compared these estimates to the masses obtained from NFW fits. We find that the distribution of mass ratios $\log_{10}(M_{\mathrm{mESR}}/M_{\mathrm{NFW}})$ has a median logarithmic offset of $0.17$~dex, indicating that imposing an NFW fit can underestimate cluster masses (see \cref{sec:enclosed_mass}). However, we note that this offset is accompanied by substantial intrinsic scatter, with a broad 16th--84th percentile range, indicating that the magnitude of the bias varies from cluster to cluster.

If not accounted for, such profile-induced systematics could propagate into the halo mass function, biasing cluster abundance and clustering analyses  \citep[e.g.][]{vikhlinin_2009, Tinker_2008}. This, in turn, affects inferred cosmological parameters, including $\sigma_8$ and $\Omega_{\rm m}$ \citep{allen_2011}. Cluster abundance measurements are also used to constrain extensions to the $\Lambda$CDM model, such as sum of the neutrino masses or the dark energy equation of state \citep{aghanim2020planck, Bocquet_2019}. These parameters affect the growth of the structure and, as such, are partially degenerate with uncertainties in the mass calibration.

Halo mass also determines the halo bias that governs the large-scale clustering of galaxy clusters. Consequently, profile-induced mass biases can propagate into clustering and cross-correlation analyses (e.g. cluster–galaxy or cluster–CMB lensing correlations).


Using a more model-agnostic approach to mass estimation also enables a self-consistent calibration at multiple overdensities. The NFW profile and other standard parametric models enforce a fixed relation between $M_{200}$, $M_{500}$ and the underlying density profile by construction. ESR allows these quantities to be derived directly from the data, allowing self-consistent but not artificially constrained mass estimates across different radial ranges.

Beyond mass calibration, enforcing a fixed parametric density profile can also bias inferences about cluster internal structure \citep{Becker_2011}. Parametric fits implicitly impose specific relations between mass, concentration, and inner density slope, potentially masking real variations in cluster profiles \citep{dutton_2014, umetsu_2020_review}. In contrast, more model-independent reconstructions, such as ESR, provide a framework for identifying departures from standard parametric forms. This departures can be identified as signatures of baryonic processes (eg. feedback, gas cooling ...) that modify the underlying dark-matter distribution. This can help in separating the contributions from baryons and dark matter, and test feedback models that make predictions for how baryons reshape dark-matter haloes.

More broadly, by not enforcing a specific dark matter model, ESR can help constrain the nature of dark matter itself. The NFW profile is derived from Cold Dark Matter (CDM) simulations, while model-independent methods can help identify non-CDM features, such as extended cores predicted by Self-Interacting Dark Matter (SIDM; \citealt{spergel2000observational, Rocha_2013}). While we do not attempt to distinguish between dark matter models in this work, we see that ESR-preferred functions seem to prefer shallower inner profiles which highlight the importance of allowing such features to be identified instead of excluded by construction.

Model-independent reconstructions may also enable the identification of other halo features not captured by NFW-like profiles, such as the splashback radius \citep{diemer_2014, more_2015}, when applied to data with sufficient radial coverage and signal-to-noise.

\subsection{Methodological caveats and extensions}
\label{sec:extensions}

\subsubsection{ESR and algorithmic improvements}

There are only two degrees of freedom in ESR. One relates to the choice of operators used in the search. Here, we have used a minimal set of basic operators: $\{x, a, \text{inv}, \exp, \log, +, \times, -, /, \text{power}\}$. One could, in principle, extend this set by including additional operators. Nevertheless, all the commonly used dark matter profiles that we tested can be constructed from this operator set. Therefore, the chosen basis is sufficient for comparing these models to alternative functional forms, even though yet-better functions may be constructed using operators that we do not consider.

Nonetheless, it is interesting to consider how ESR could be extended to reach higher complexities. A fully exhaustive search at these levels is computationally prohibitive with the current algorithm. Instead, one possible approach is to use a stochastic search, seeded with the functions found by ESR. Alternatively, a deterministic approach could be used to systematically explore functions in the neighbourhood of those already found by ESR. By systematically adding operators to the high-performing functions already identified, this approach focuses on a targeted region of function space that is more likely to contain well-fitting profiles. In this way, the exploration remains ``exhaustive'' within this relevant region, as all reasonable extensions of the best functions are considered, rather than relying on purely random sampling. 

Regarding parameter optimisation, we employed a combination of optimisation methods to balance accuracy and computational cost. For functions above complexity $6$, we first optimised all candidates for a small subset of clusters, identified the best-performing expressions, and then optimised only this refined set across the full sample. In other words, we targeted the most promising functions rather than exhaustively optimising every candidate. As detailed in Section $\ref{optimisation}$, we verified that this strategy was sufficient to identify the best-fitting profiles by confirming that the selected functions included a large number of high-performing models, and that those excluded were suboptimal. To ensure robustness, we repeated optimisations using multiple initial conditions and confirmed that the global minimum was consistently recovered.

Other parameter-optimisation strategies could also be incorporated into ESR in future work. For example, global optimisation algorithms such as Differential Evolution \citep{StornPrice1997}, Simulated Annealing \citep{Kirkpatrick1983}, or Particle Swarm Optimisation \citep{KennedyEberhart1995} provide efficient ways of exploring multi-modal likelihood surfaces and could complement the current optimisation scheme.


\subsubsection{Extensions and systematics in cluster analysis} \label{sec:systematics_clusters}

Another important consideration is the treatment of systematics in the data. In this work we effectively model the total mass profile within the radial range probed, implicitly assuming that the baryonic contribution is small compared to the dark matter component. This is justified for cluster-scale weak-lensing analyses, where the baryonic mass—arising primarily from the BCG and the intracluster medium—is typically subdominant at radii $R \gtrsim 100$ kpc, contributing around $1/6$th of the total mass. \citep{umetsu_cluster-galaxy_2020}. Nevertheless, baryons can be modelled explicitly using auxiliary data \citep{beauchesne2025comprehensiveseparationdarkmatter, Allingham_2024} and added to the modelling scheme. An alternative option to this is simply to interpret the ESR-derived profiles as total mass profiles, without attempting to separate dark matter and baryons.  

Similarly, we have assumed that the two-halo term can be ignored at the scales studied following \citet{umetsu_weak_2020}. They found the two-halo term's contribution to the stacked ESD profile to be negligible across their entire radial range (see their Fig. $4$). Even at their outermost datapoint ($R \approx 3 \, h^{-1} \text{Mpc}$), the two-halo term remained roughly an order of magnitude smaller than the fitted one-halo component. Since our analysis uses the same data and reaches comparable scales, this approximation is well justified.

Beyond this empirical motivation, there is also a more fundamental methodological consideration: standard implementation of the two halo-term assume a \(\Lambda\)CDM population with an NFW density profile and a mass--concentration relation calibrated on simulations. Given that the goal of this work is to infer dark matter density profiles directly from the data, adopting a two-halo term computed under NFW assumptions would be inconsistent. In principle, one could compute a self-consistent two-halo term for each candidate profile, but this would be computationally expensive. In the near-term, the most practical option would be to test the stability of the preferred profiles against the radial range of the data (e.g., by re-fitting with the outermost bins removed).


Our model also relies on two other standard simplifying assumptions that could be expanded upon in future work. First, we assume spherical symmetry for the clusters. Incorporating triaxiality as a free parameter within ESR could provide a more realistic description of cluster shapes. Second, we assume that the cluster centre is known. In practice, the true centre can be uncertain, and misentering can affect the inferred inner profile. In the southern XXL field, where a BCG catalogue is available, we tested centring on the BCG instead of the X-ray peak and found that this did not remove the dips observed towards the centre of the ESD profiles in many clusters.
ESR could naturally incorporate the cluster centre as an additional parameter, constrained by informative priors from X-ray, optical, or SZ observations. Relaxing these assumptions would provide a more flexible modelling framework and can improve the fidelity of the inferred profiles when higher quality-data becomes available.


\subsection{Comparison to literature}


There have been other attempts to discover density profiles without reliance on the NFW model. In the observational domain, several studies have tried to recover mass density profiles directly from data. For example, a number of works have focused on moving beyond fixed analytic profiles by using non-parametric or semi-parametric techniques. These include the free-form radial reconstructions \citep[e.g.,][]{johnston2007crosscor, Mistele_2024}, which move beyond fixed analytic assumptions by allowing the data to determine the profile shape. Interestingly, such reconstructions often give enclosed-mass estimates that are close to those obtained from standard NFW fits \citep{Mistele_2025}. We find a similar behaviour here, with ESR masses being on average slightly higher than NFW-derived masses. That said, our analysis is sufficiently precise to reveal clearly that NFW is not the best profile.


A key advantage of these non-parametric methods is their computational efficiency which makes them significantly cheaper to run than the exhaustive search performed by ESR. However, while they minimise modelling assumptions, they result in binned numerical estimates rather than closed-form analytic expressions. As such, they lack the ability to extrapolate the density profile beyond the radial range of the data, which is a unique advantage of the symbolic regression approach presented here. Furthermore, these methods often require specific binning choices or regularisation schemes that can be difficult to interpret physically, and it would be difficult to integrate them into cosmological analyses which assume analytic forms for cluster halo density profiles. Having a functional form is clearly useful for summarising the dark matter distribution within a larger analysis.

Beyond these profile-modelling efforts, other work has focused on cluster mass calibration by bypassing the density profile entirely and learning a direct mapping from weak-lensing maps to halo mass. For instance, \citet{Gupta2018} used Convolutional Neural Networks (CNNs) to infer cluster masses from weak-lensing maps, achieving high predictive accuracy. However, these ``black box'' approaches function as non-linear calibrators for the specific simulation physics they were trained on. They implicitly assume the halo structure follows the standard $\Lambda$CDM predictions rather than testing the validity of those predictions. While they excel at parameter estimation (e.g., $M_{200}$), they do not provide a mathematical description of the halo structure itself.

Finally, a number of recent studies have focused on learning the mapping between haloes and their density profiles using $N$-body simulations. These approaches aim to link the final density field to the haloes' evolutionary histories. For example, \citet{Lucie_Smith_2022_discovering,Lucie_Smith_2022} used deep neural networks and autoencoders to compress latent representations of density profiles. This approach performs a dimensionality reduction trying to find the minimal latent components to describe a halo. This is conceptually similar to minimising an information criterion, as we do with the MDL metric, though it primarily optimises the number of describing variables rather than the complexity of the functional form itself. Building on this, \citet{lucie-smith_explaining_2024} investigated which features of the mass accretion history most strongly influence the final shape of the profile. However, while a trained neural network technically constitutes a closed-form mathematical function (composed of matrix operations), it remains a ``black box'' containing thousands of parameters. Unlike the symbolic expressions we derive, these networks do not output simple, interpretable analytic formulae. 

Within the specific context of symbolic regression, \citet{Thing_2025} recently presented a comprehensive benchmark of symbolic regression algorithms applied to synthetic cosmological datasets, including dark matter halo profiles. Their results highlighted that while data-driven algorithms can successfully recover profile shapes (such as NFW or cored models) from high-precision data, they are sensitive to noise and data quality. Our analysis differs from this in both scope and application. While \citet{Thing_2025} assessed the ability of heuristic algorithms to locate the NFW functional form within a vast search space, we use an exhaustive search (ESR) on observational data. Since ESR explicitly evaluates every function up to complexity $10$, the NFW profile is by definition included as a possible model. Consequently, the question is not whether the algorithm succeeds in considering and evaluating NFW (an issue of reliability of the algorithm), but rather whether or not NFW scores highly (an issue with the data and its error model). Our finding that NFW is not the preferred model represents a robust statistical rejection based on the data, rather than a failure of the search algorithm to locate the function.

\subsection{Future applications and datasets}


A natural next step is to apply this method to upcoming weak-lensing datasets that offer higher signal-to-noise ratios or larger sample sizes. Surveys such as the Dark Energy Survey \citep[DES;][]{abbott2018}, \textit{Euclid} \citep{laureijs2011} or CLASH \citep{Postman2012CLASH}, could provide more stringent constraints on halo profiles and allow a direct comparison with the HSC–XXL results. Such data would also test whether the functional forms identified here are stable across different cluster samples observed with different instruments and depth.

There is no need to limit ourselves to data coming from WL. The framework presented here can be applied to any dataset that constrains the radial distribution of dark matter in galaxies or galaxy clusters. The core idea is simple: use symbolic regression to generate candidate expressions for the density profile $\rho(r)$, compute the corresponding observables and rank the models using the MDL principle. 



Kinematic data, such as galactic rotation curves from the SPARC database \citep{Lelli_2016} or integral-field spectroscopy from MaNGA \citep{bundy2015}, would provide powerful complementary constraints. These data are highly sensitive to the small galactic radii ($r \lesssim 0.1 \text{ Mpc}$) that WL cannot constrain. This is particularly relevant to our finding that the HSC data's inner profile is unconstrained due to large ESD errors. Other multi-wavelength probes, such as X-ray or SZ observations of the intracluster medium, could also be studied independently.

Moreover, combining different datasets is straightforward. ESR naturally accounts for differences in the number of data points and measurement precision; one simply needs to combine the description lengths from each probe, in a similar way to what was done here for multiple galaxy clusters, to perform a joint analysis.

Another interesting direction is to study profile evolution with redshift. By applying ESR to cluster samples split into redshift bins, one could test whether the preferred functional forms evolve with cosmic time.

Finally, ESR can be directly applied to simulations. Since NFW itself comes from $N$-body simulations, it would be interesting to test whether ESR recovers NFW or identifies alternative functional forms already in simulated data and to compare these to the profiles preferred by observations. Such a comparison could also help understand where baryonic physics or other effects may drive differences between dark matter-only simulations and reality.

\section{Conclusions} \label{section:conclusions}

The NFW profile has long served as the standard model for describing dark matter halos, but it remains a phenomenological fit to simulation data rather than a theory-driven prediction or empirically favoured function. In this work, we apply Exhaustive Symbolic Regression (ESR) to derive dark matter density profiles directly from observations. In particular, we apply the method to the HSC–XXL dataset of 149 galaxy clusters. This enables a data-driven exploration of possible halo profiles without imposing a fixed parametric form. Our conclusions are as follows:

\begin{itemize}
    \item ESR discovers functions that statistically outperform common dark matter profiles (e.g., NFW, Einasto, Burkert) in terms of both accuracy and simplicity. Notably, the data favour functions with only two free parameters. Profiles requiring more parameters are generally penalised by the MDL metric, suggesting that the current weak lensing data cannot robustly constrain more than two structural parameters.  We specifically highlight two analytically integrable profiles that outperform NFW: the exponential decay with a central hole given in \cref{eq:best_integrable} and the cuspy profile presented in \cref{eq:second_integrable}.
    \item The weak lensing data we use primarily constrain the outer halo, leaving the inner profile poorly constrained. Due to the weak-lensing shape noise decaying roughly as $1/R$ for radial bins, the data naturally prefer fitting functions that fit well on the outer range of the data.
    \item Despite the limited inner sensitivity, ESR consistently selects profiles with shallow central behaviour and a maximum in the ESD within the range of the data. The radius at which the maximum appears varies across clusters, shifting inward for high-SNR systems and outward for lower SNR clusters. For these low-SNR systems, the inner measurements are consistent with zero, so any non-trivial features in the ESD are only required at larger radii.

    \item We used the best-fitting ESR models to calculate the enclosed mass of the clusters ($M_{200}$). These are inferred by averaging over the ensemble of high-performing ESR functions and give physically reasonable values that are largely consistent with standard NFW results although a bit higher on average.
    \item We used the best-fitting ESR functions to infer cluster masses ($M_{200}$) by marginalising over profile shape. These ESR-derived masses are broadly consistent in trend with NFW estimates, but are systematically higher on average. The median logarithmic mass ratio is $0.17$\,dex, which would imply that NFW underestimates masses by $\approx 48\%$ relative to ESR. Although the distribution is broad (16th--84th percentile range of $[-0.52,\,1.10]$\,dex), this shows that using the wrong profile can introduce bias in mass inferences. 
    \item We tested whether allowing different clusters to adopt different functional is statistically preferred to having a universal profile. We tried both allowing for a different profile per-cluster and allowing each cluster to choose between just two profiles. 
    Allowing multiple
    models performs worse in description length than a single universal profile due to the penalties associated with specifying multiple functions and encoding their assignments. This is consistent with the expected self-similarity of cluster haloes.
    \item The ESR framework is readily applicable to any other dataset that constrains the full distribution of matter, including constraints from galaxy kinematics, X-ray observables, the thermal SZ effect and multi-wavelength combinations. Studying these complementary probes would help better constrain the inner regions of clusters and galaxies and test whether the structural features that we see here remain preferred.
    \item Our results depend on the maximum complexity of the functions considered. Here, we have used ESR to complexity $10$. Future developments in symbolic regression could extend this search to higher complexities, helping to determine whether the features identified here persist as the optimal profiles or whether the true minimum of the description length (maximum of the posterior probability) lies at higher complexity.
\end{itemize}

ESR provides a flexible and fully data-driven framework for deriving, comparing, and interpreting dark matter halo profiles. By replacing fixed analytic assumptions with a systematic function search governed by MDL, it offers a powerful approach for connecting observations to empirical models of dark matter structure.

\section*{Acknowledgements}

We thank Gary Mamon, Richard Stiskalek and Keiichi Umetsu for useful discussions.

AM acknowledges support financial support for a DPhil studentship from the Department of Physics at Oxford. TY acknowledges the support of a UKRI Frontiers Research Grant [EP/X026639/1], which was selected by the
European Research Council. 
DJB was supported by the Simons Collaboration on ``Learning the Universe'' and is supported by Schmidt Sciences through The Eric and Wendy Schmidt AI in Science Fellowship.
HD is supported by a Royal Society University Research Fellowship (grant no. 211046).
PGF acknowledges support from STFC and the Beecroft Trust. 

 This work used the Glamdring cluster at the University of Oxford. Additionally, we used resources provided by the Cambridge Service for Data Driven Discovery (CSD3) operated by the University of Cambridge Research Computing Service (www.csd3.cam.ac.uk), provided by Dell EMC and Intel using Tier-2 funding from the Engineering and Physical Sciences Research Council (capital grant EP/P020259/1), and DiRAC funding from the Science and Technology Facilities Council (www.dirac.ac.uk).

\section*{Data Availability}

The data underlying this article will be shared on reasonable request to the corresponding author.


\bibliographystyle{mnras}
\bibliography{refs} 

\bsp	
\label{lastpage}
\end{document}